\documentclass[fleqn,usenatbib]{mnras}
\usepackage{amsmath}
\usepackage{graphicx}
\usepackage{newtxtext}
\usepackage{newtxmath}
\title[Formation of Mass-gap BHs from NS XRBs with Super-Eddington Accretion]{Formation of Mass-gap Black Holes from Neutron Star X-ray Binaries with Super-Eddington Accretion}

\author[S.-J. Gao, X.-D. Li and Y. Shao]{
	Shi-Jie Gao,$^{1,~2}$
	Xiang-Dong Li$^{1,~2}$\thanks{E-mail: lixd@nju.edu.cn}, Yong Shao$^{1,~2}$
	\\
	$^1$School of Astronomy and Space Science, Nanjing University, Nanjing, 210023, People's Republic of China\\
	$^2$Key Laboratory of Modern Astronomy and Astrophysics, Nanjing University, Ministry of Education, Nanjing, 210023, People's Republic of China
}

\date{Accepted 2022 May 19. Received 2022 May 19; in original form 2022 January 5}
\pubyear{\today}

\begin{document}
\label{firstpage}
\pagerange{\pageref{firstpage}--\pageref{lastpage}}
\maketitle

\begin{abstract}
Electromagnetic and gravitational wave observations indicate that there is dearth of compact objects with mass $\sim 2.5-5~{\rm M_\odot}$. This so-called ``mass gap'' may be linked to the supernova explosion mechanisms that produce neutron stars (NSs) and black holes (BHs). However, the existence of a few mass-gap compact objects, some of which have been confirmed to be BHs, poses a challenge to the traditional theory of black hole formation. In this work we investigate the possible formation channel of BHs from accretion-induced collapse (AIC) of NSs in X-ray binaries. In particular, we consider the influence of super-Eddington accretion of NSs. Recent observations of ultraluminous X-ray pulsars suggest that their apparent luminosities may reflect the true accretion luminosities of the accreting NSs, even exceeding the Eddington limit by a factor of $\gtrsim 100$. Thus, NSs accreting at a super-Eddington accretion rate may rapidly grow into BHs in intermediate/low-mass X-ray binaries. Based on the super-Eddington accretion disk models, we have investigated the evolution of NSs in intermediate/low-mass X-ray binaries by combining binary population synthesis and detailed stellar evolutionary calculations. We show that super-Eddington accretion plays a critical role in mass growth of NSs, and the final masses of the descendant BHs are heavily dependent on the NS magnetic fields, the metallicity of the donor star, and the bifurcation period of the binaries. AIC of NSs may account for some of the observed mass-gap BHs like GRO~J0422+32. We also present the parameter distributions of the potential mass-gap BHs in a Milky Way-like galaxy, and point out that future space-based gravitational wave observations may provide important test of or constraints on the formation of mass-gap BHs from the AIC channel.
\end{abstract}


\maketitle

\section{Introduction} \label{section:introduction}

\quad Compact objects (COs) including white dwarfs (WDs), neutron stars (NSs) and black holes (BHs) are the end products of stellar evolution. Considering the initial mass function of stars in the Galaxy, it was traditionally expected a continuous mass function covering all types of COs. However, this intuition seems to be inconsistent with the fact that there is dearth of observed COs with mass $\sim 2.5-5~{\rm M_\odot}$ \citep{Bailyn+1998,Ozel+2010,Farr+2011}. Recent electromagnetic observations of Galactic NSs suggest a likely maximum mass of $\sim 2.0-2.3~{\rm M_{\odot}}$. For example, \cite{Cromartie+2020} measured the mass of the millisecond pulsar J0740+6620 to be $2.14_{-0.18}^{+0.20}~{\rm M_\odot}$ using relativistic Shapiro delay observations, which was later updated to be $2.03^{+0.10}_{-0.08}~{\rm M_\odot}$ \citep{Farr+2020}. \cite{Kandel+2020} reported a black widow binary system PSR~J1959+2048 with the NS mass $2.18\pm0.09~{\rm M_\odot}$, and a redback binary system PSR~J2215+5135 with the NS mass $2.28^{+0.10}_{-0.09}~{\rm M_\odot}$ (see also \citealt{Linares+2018}). On the other hand, the minimal mass of BHs in X-ray binaries (XRBs) was inferred to be $\sim 5~{\rm M_\odot}$ \citep{Casares+2014}.

The origin of the mass gap\footnote{Here the `mass gap' refers to the lower mass gap ($\sim 2.5-5~{\rm M_{\odot}}$) between the most massive NSs and the lightest BHs rather than the (upper) pair-instability mass gap for BHs ($\sim 64-161~{\rm M_\odot}$, e.g., \citealt{Woosley+2021}).} may be related to the supernova (SN) explosion mechanisms. For example, the rapid and delayed SN models in \cite{Fryer+2012} differ in the growth timescales of the instabilities driving the explosion of massive stars. While the rapid SN model can reproduce the mass gap, the delayed model predicts the mass gap to be populated \citep{Belczynski+2012}.

However, recent multi-messenger observations indicate that the mass gap is at least partly populated. According to \cite{Kreidberg+2012}, the BHs in two XRBs (4U1543-47 and GRO~J0422+32) may be less massive than $\sim 4~{\rm M_\odot}$, suggesting that the absence of low-mass BHs could be due to a potential observational artifact caused by the systematic uncertainties affecting ellipsoidal fits and hence inclination measurements. Microlensing observations also suggest that the mass distribution of dark lenses is consistent with a continuous distribution of stellar-remnants, and a mass gap between the NS and BH masses is not favored (\citealt{Wyrzykowski+2020}, see however \citealt{Mroz+2021}). Using the radial velocity and photometric variability data, \cite{Thompson+2019} reported the discovery of a non-interacting low-mass BH (or an unexpectedly massive NS) with mass $3.3_{-0.7}^{+2.8}~{\rm M_\odot}$ accompanied by a giant star (2MASS~J05215658+4359220). Observations of the gravitational wave (GW) event GW190814 revealed a mass-gap CO of mass $2.59_{-0.09}^{+0.08}~{\rm M_\odot}$ and a BH of mass $\sim 23~{\rm M_\odot}$ \citep{Abbott+2020}. Most recently, \cite{Abbott+2021} reported another mass-gap CO with mass $2.83_{-0.42}^{+0.47}~{\rm M_\odot}$ in GW200210\_092254 during the third observing run of the advanced \textit{LIGO} and the advanced \textit{Virgo}. Although the existence of the mass gap is still controversial, it is commonly accepted that there are relatively few low-mass ($\lesssim5~{\rm M_\odot}$) BHs.

In addition to the stellar collapse channel, NSs and BHs can form from accretion-induced collapse (AIC) and merger-induced collapse in interacting binaries \cite[e.g.,][]{Ivanova+2008,Ablimit+2021}. In the AIC channel, a WD/NS accretes material from its donor star and grows in mass until a critical mass is reached, when the degenerate pressure cannot support the star against gravity, and the WD/NS collapses to be an NS/BH \citep{Nomoto+1991,Timmes+1996,Ivanova+2004,Wang+2020RAA}. Similar processes may take place in the active galactic nucleus disks where NSs/WDs embed \citep[e.g.,][]{Perna+2021,ZhuJinPing+2021}. In this work, we focus on the evolution of isolated interacting binaries consisting of an NS and the related AIC process. Since a $1.4~{\rm M_\odot}$ NS needs to accrete at least $\sim 1~{\rm M_\odot}$ to collapse to a be BH, AIC can only take place in intermediate/low-mass X-ray binaries (I/LMXBs), in which the donor mass $\lesssim 6~{\rm M_\odot}$ and mass transfer proceeds via Roche-lobe overflow (RLOF) on a timescale $\sim 10^{7}-10^9~{\rm yr}$. In high-mass X-ray binaries (HMXBs) in which the donor mass $\gtrsim 10~{\rm M_\odot}$, the 
mass transfer is dynamically unstable with a duration too short to allow efficient mass accretion \citep{Tauris+2006}.

There are $\sim 200$ LMXBs in the Galaxy and their orbital periods are relatively short, typically $\lesssim10~{\rm days}$ \citep{Liu+2007}. This means that most LMXBs have experienced common envelope evolution \citep[CEE, ][]{Paczynski+1976,Ivanova+2013}. A more massive secondary star (i.e., the donor star) usually has more kinetic energy to expel the envelope of the primary star (i.e., the progenitor of the NS) during the spiral-in process, and is more likely to survive CEE. Thus, IMXBs have a much higher birthrate than LMXBs, and they will become LMXBs when the donor mass decreases to be $\lesssim 1~{\rm M_\odot}$. This has led to the suggestion that most LMXBs have actually evolved from IMXBs \citep{Pfahl+2003}.

The accretion rate for an NS in I/LMXBs is conventionally thought to be limited by the Eddington accretion rate ($\dot M_{\rm Edd}\simeq 4\times10^{-8}~{\rm M_\odot~yr^{-1}}$ for a $1.4~{\rm M_\odot}$ NS). However, super-Eddington accretion has been suggested both observationally and theoretically.
Ultraluminous X-ray sources \citep[ULXs; for recent reviews, see][]{Kaaret+2017,Fabrika+2021} are off-nuclear point-like X-ray sources with inferred isotropic luminosities $L_{\rm X}\gtrsim 10^{39}~{\rm erg~s^{-1}}$, which significantly exceed the Eddington limit luminosity ($\sim 2\times10^{38}~{\rm erg~s^{-1}}$) for a $1.4~{\rm M_\odot}$ NS. The discovery of pulsations in ULX M82~X-2 indicates that it is powered by an accreting NS rather than a BH \citep{Bachetti+2014}, and its X-ray luminosity can reach $\sim 2\times10^{40}~{\rm erg~s^{-1}}$. Subsequently, quite a few ULX pulsars have been discovered, including NGC5907~ULX-1 \citep{Israel+2017+NGC5907ulx1}, M51~ULX-7 \citep{Rodrguez+2020}, NGC7793~P13 \citep{Furst+2016,Furst+2018,Israel+2017+NGC7793P13}, NGC300~ULX-1 \citep{Carpano+2018}, SMC~X-3 \citep{Tsygankov+2016,Townsend+2017}, NGC2403 ULX \citep{Trudolyubov+2007}, Swift~J0234.6+6124 \citep{Doroshenko+2018,vandenEijnden+2018} and RX~J0209.6-7427 \citep{Chandra+2020}.
The reasons why an NS can produce such a high luminosity are still under debate. Some authors \citep{King+2017,Middleton+2017,King+2019,King+2020} suggest that the ULX radiation is strongly anisotropic and the apparent luminosity $L_{\rm app}$ is amplified with a beaming factor $\Tilde b\ll 1$, while others \citep{Dall+2015,Eksi+2015,Mushtukov+2015,Mushtukov+2017,Mushtukov+2019,Chashkina+2019} argue that the apparent luminosity of the ULX pulsars should be close to their true luminosity (i.e., $\Tilde b\lesssim 1$), because the observed sinusoidal pulse profiles in M82 X-2, NGC5907 ULX-1 and NGC77893 P13 do not favor a strong beaming. \cite{Mushtukov+2021} used Monte--Carlo simulations to trace the photon emission from the NS ULXs and pointed out that strong beaming is inconsistent with the observations of a large pulsed fraction.

\cite{Kuranov+2020} investigated the properties of NS ULXs with a binary population synthesis (BPS) method, and showed that the observed populations of NS ULXs in spiral galaxies can be explained without assuming any strong beaming effects. In this work, we explore under what circumstance can super-Eddington accretion produce mass-gap BHs in I/LMXBs and the properties of the descending BH binaries. This paper is organized as follows. We describe our method of BPS and detailed evolutionary calculations of I/LMXBs in Section~\ref{section:method}. In Section~\ref{section:results}, we demonstrate the properties of the BH binaries from the combination of BPS and evolutionary tracks. Finally, we discuss the uncertainties in our work and give a brief summary in Section~\ref{section:dis&sum}.

\section{Methods and Calculations}\label{section:method}

\quad We use the BPS code \textsc{BSE} developed by \cite{Hurley+2002+BSE} and modified by \cite{Kiel+2006} and \cite{Shao+2014+Be} to follow the formation of incipient NS I/LMXBs. We briefly list the input parameters and the key factors in our model as follows.

We use a Monte--Carlo method to evolve $10^8$ primordial binaries, assuming a constant star formation rate $5~{\rm M_\odot~yr^{-1}}$ \citep{Smith+1978} in the past $12~{\rm Gyr}$. For the initial parameters of the binaries, we randomly draw them from their independent probability distribution functions. The initial mass $M_1$ of the primary is distributed in the range of $3-100~{\rm M_\odot}$, following the initial mass function of \cite{Kroupa+1993}. We assume a uniform distribution of the mass ratio between the secondary and primary masses $q=M_2/M_1$ in the range of $0-1$. For the metallicity we consider both $Z=0.02$ (case~1) and $0.001$ (case~2). The initial orbital separation $a_0$ is assumed to follow a logarithmically uniform distribution in the range of $3~{\rm R_\odot}-10^4~{\rm R_\odot}$. We assume the eccentricity of all the primordial binaries to be zero, as its effect on the evolution of I/LMXBs is minor (\citealt{Hurley+2002+BSE}).

We use the numerically calculated critical mass ratio $q_{\rm c}$ in \cite{Shao+2014+Be} to determine whether the mass transfer in the primordial binaries is dynamically stable, and adopt the values of $q_{\rm c}$ in the rotation-dependent model which can reproduce the observed properties of Be/XRBs \citep[for a review of Be/XRBs, see][]{Reig+2011}. In this model mass transfer becomes highly non-conservative when the accreting secondary reaches critical rotation.

If the mass transfer is unstable, we use the energy conversation equation \citep{Webbink+1984} to treat the CEE. We take $\alpha_{\rm CE}=1.0$, where $\alpha_{\rm CE}$ is the CEE efficiency defined as the fraction of the available orbital energy that can be used to eject the common envelope. For the binding energy parameter $\lambda$ of the primary's envelope, we use the values calculated by \cite{Xu+2010} and modified by \cite{Wang+2016} for a range of massive and intermediate-mass stars under different evolutionary stages.

We adopt the criterion in \citet{Fryer+2012} to discriminate core-collapse supernovae (CCSNe) from electron-capture supernovae (ECSNe), and their rapid supernova mechanism for the NS formation. The natal kick velocity $v_{\rm k}$ of newborn NSs obeys a Maxwellian distribution with the dispersion $\sigma_{\rm k}=265~{\rm km~s^{-1}}$ \citep{Hobbs+2005} and $\sigma_{\rm k}=50~{\rm km~s^{-1}}$ \citep{Dessart+2006} for CCSNe and ECSNe, respectively.
\begin{figure*}
	\centering
	\includegraphics[width=0.8\linewidth]{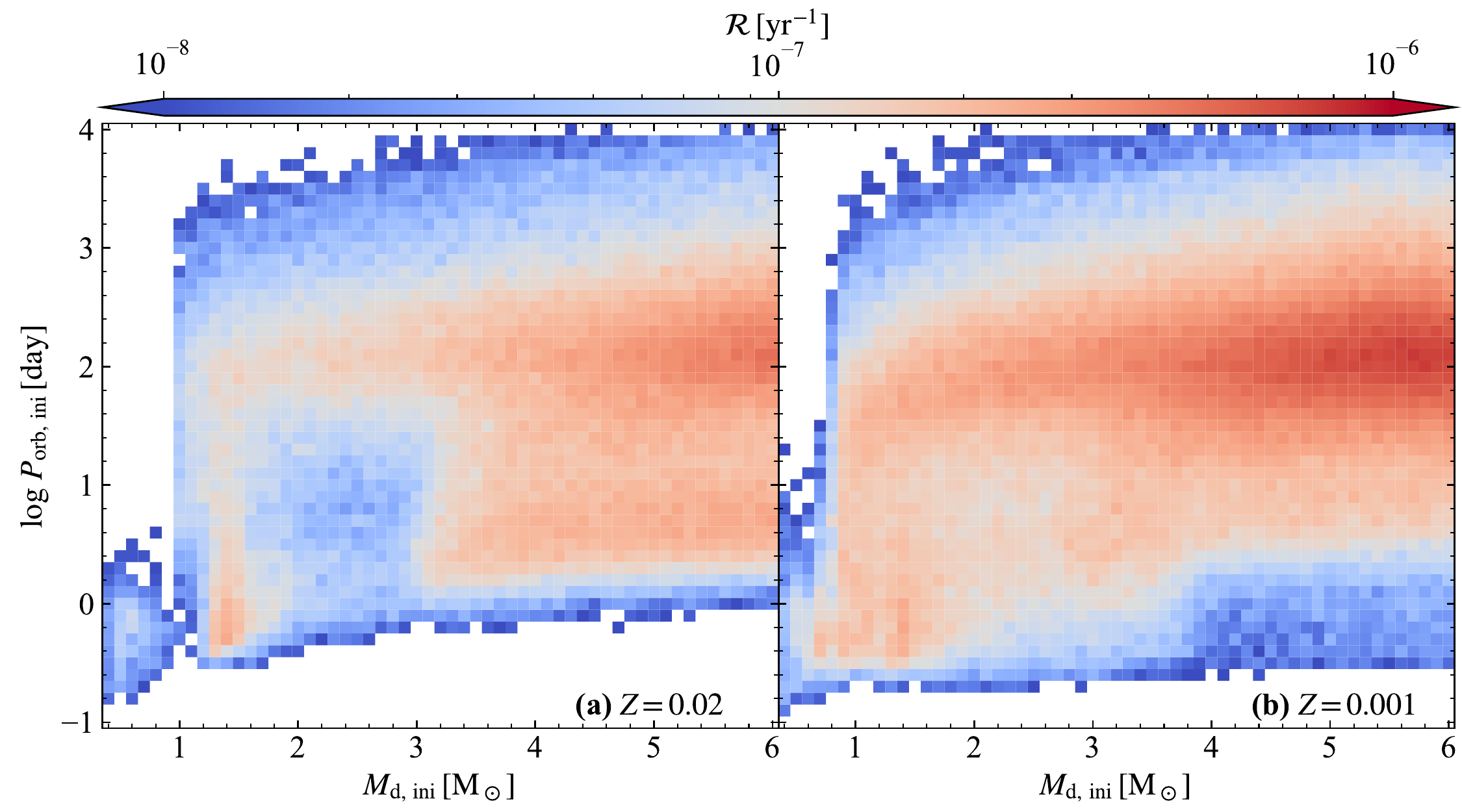}
	\caption{
		The birthrate distribution of incipient I/LMXBs in the initial donor star mass-orbital period ($M_{\rm d,~ini}$-$P_{\rm orb,~ini}$) plane. Panels (a) and (b) correspond to cases~1 and 2. respectively.
		\label{fig:birthrate2d}}
\end{figure*}

\begin{figure*}
	\centering
	\includegraphics[width=\linewidth]{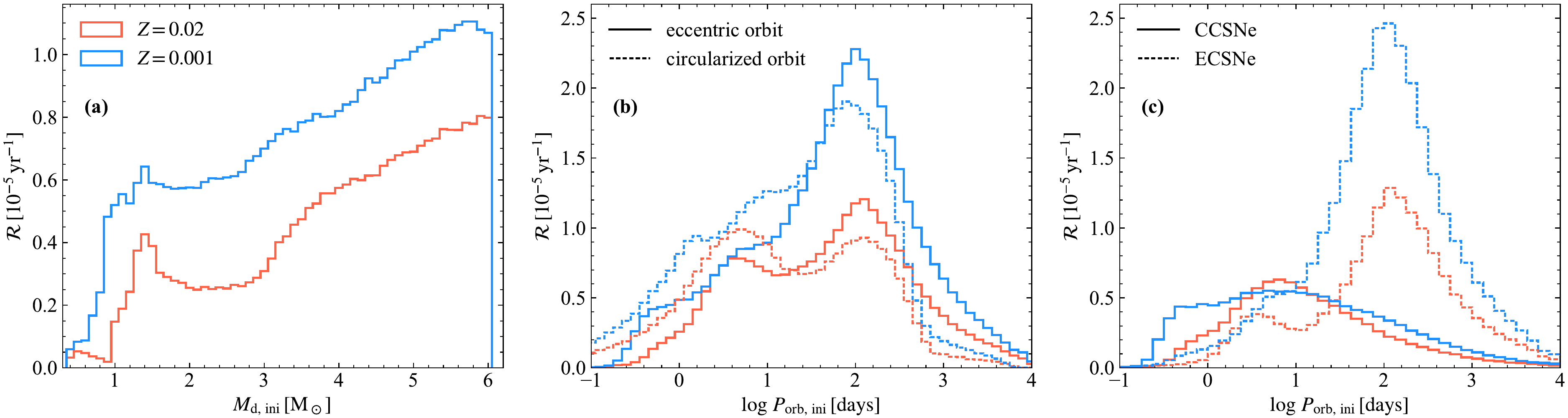}
	\caption{
		The birthrate distribution of incipient I/LMXBs as a function of the donor star mass (Panel~a) and the orbital period (Panel~b and c). The blue and red lines show the results in cases~1 and 2, respectively. The solid and dashed lines in panels~(b) and (c) represent the birthrate distribution as a function of the orbital period for eccentric and circularized orbits, and for NSs formed from CCSNe and ECSNe, respectively.
		\label{fig:birthrate1d}}
\end{figure*}

\subsection{Formation of incipient NS I/LMXBs}\label{subsection:formation}

\quad We select the incipient NS I/LMXBs with the periods $P_{\rm orb,~ini}<10^4~{\rm days}$ and the donor/companion mass $M_{\rm d,~ini}<6~{\rm M_\odot}$ at the birth of the NS.
The total birthrates of the incipient NS I/LMXBs are estimated to be $\sim 2.46\times10^{-4}~{\rm yr^{-1}}$ and $4.41\times10^{-4}~{\rm yr^{-1}}$ for cases~1 and 2, respectively. Low metallicity stars have shorter nuclear lifetime which allows them to have more chances to interact with their companions in a relatively wide orbit \citep{Hurley+2002+BSE}. In addition, they suffer less wind mass loss, leading to less increase in orbital separation previous to CEE, which also makes the binaries more likely to interact in a closer orbit \citep{Kiel+2006}. Thus, the birthrate of the NS I/LMXBs in case~2 is slightly higher than in case~1. In \autoref{fig:birthrate2d}, we use different colors to represent the birthrate of incipient I/LMXBs in the the donor mass-orbital period ($M_{\rm d,~ini}$-$P_{\rm orb,~ini}$) plane. There is almost no binary with $M_{\rm d,~ini}\lesssim 1~{\rm M_\odot}$ and $P_{\rm orb,~ini}\gtrsim 1~{\rm days}$ for the reason that neither magnetic braking nor the stellar evolution can trigger RLOF within $12~{\rm Gyr}$.

\autoref{fig:birthrate1d} demonstrates the birthrate of the incipient I/LMXBs as a function of the donor mass $M_{\rm d,~ini}$ (panel~a) and the orbital period $P_{\rm orb,~ini}$ (panels b and c). The red and blue curves correspond to cases~1 and 2, respectively. Panel~(a) shows that the birthrate of incipient I/LMXBs generally increases with the donor mass, because binaries with lower-mass companions are more likely to merge during the CEE \citep{Pfahl+2003}. After the SN explosions the binary orbits become eccentric because of mass loss and the kicks imparted on the NSs. We assume that orbits are quickly circularized due to tidal torques and in this process the orbital angular moment is conserved. The circularized orbital period $P_{\rm orb,~cir}=P_{\rm orb,~ecc}(1-e^2)^{3/2}$, where $P_{\rm orb,~ecc}$ and $e$ are the orbital period and eccentricity after the SN explosion, respectively. In panel~(b), the solid and dashed lines represent the eccentric and circularized orbits, respectively. The majority of incipient NS I/LMXBs are distributed in the range of $1-1000~{\rm days}$. Panel~(b) demonstrates that the birthrate in case~1 reveals a bimodal distribution peaked at $\sim 8~{\rm days}$ and $\sim 150~{\rm days}$. The origin of this bimodal feature comes from the fact that NSs formed from CCSNe and ECSNe are distributed in relatively short and long orbital period binaries\footnote{Otherwise the mass transfer would occur at the early evolutionary stage and lead to the formation of WDs instead of NSs.}, respectively (see panel c). In case~2, the bimodal feature is not as clear as in case~1, because the birthrate distribution is dominated by NSs from ECSNe in wide binaries.

\subsection{Evolution of NS I/LMXBs}\label{subsection:evolutionofILMXBs}

\begin{figure*}
	\centering
	\includegraphics[width=\linewidth]{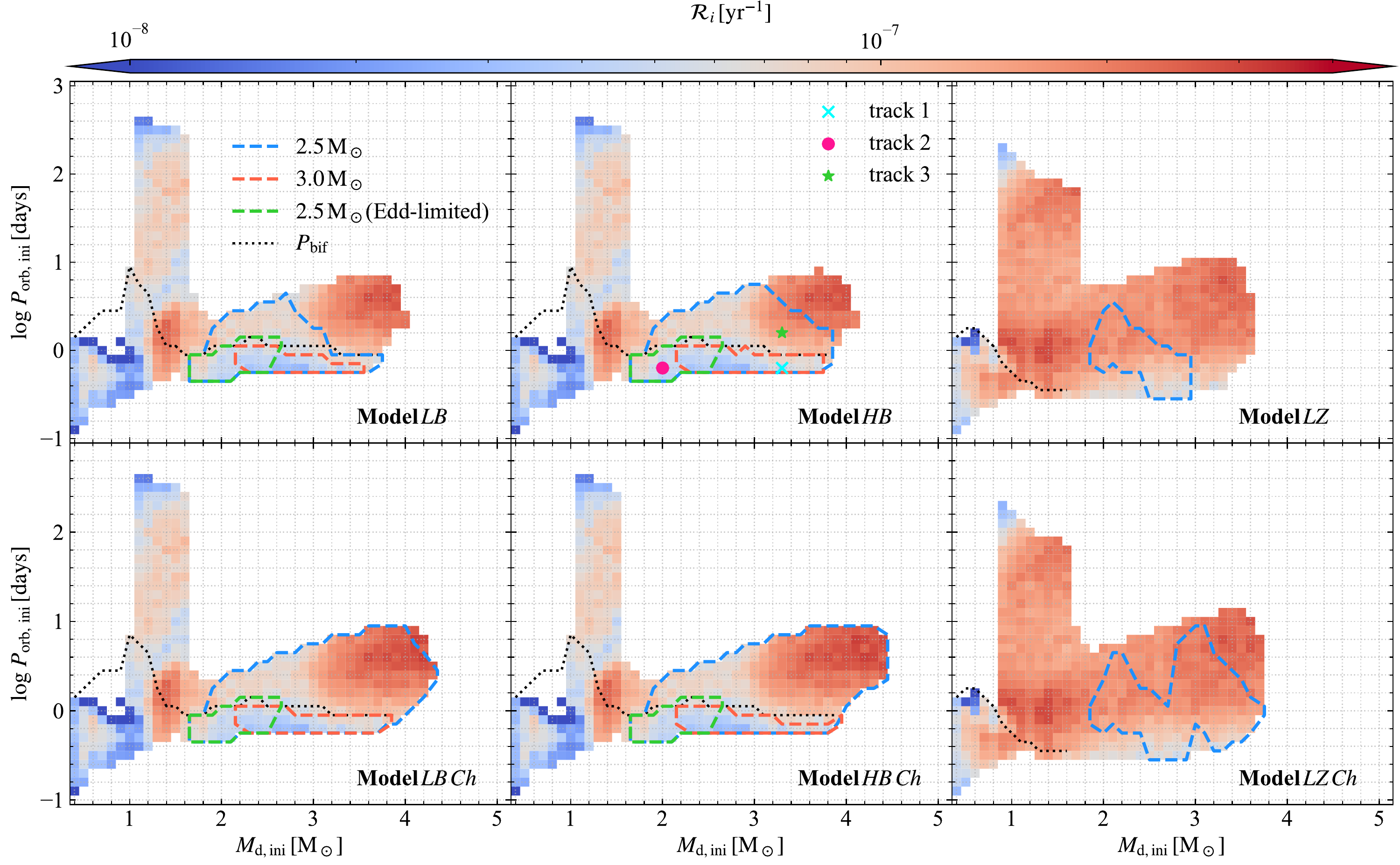}
	\caption{Allowed parameter space for stable mass transfer in the initial donor mass-orbital period ($M_{\rm d,~ini}$-$P_{\rm orb,~ini}$) plane in different models. The color in each pixel indicates the birthrate of specific I/LMXBs inferred from the BPS results. The blue and the red dashed contours indicate the final masses of the accretors $M_{\rm acc,~f}=2.5~{\rm M_\odot}$ and $3.0~{\rm M_\odot}$ respectively, and the green dashed contours indicate $M_{\rm acc,~f}=2.5~{\rm M_\odot}$ with the Eddington-limited accretion prescription for comparison. The black dotted line in each panel shows the bifurcation period as a function of the initial donor star mass. The cyan, pink and green markers in the upper middle panel show the initial positions of tracks~1, 2 and 3 evolution.\label{fig:stablemt}}
\end{figure*}

\begin{figure*}
	\centering
	\includegraphics[width=\linewidth]{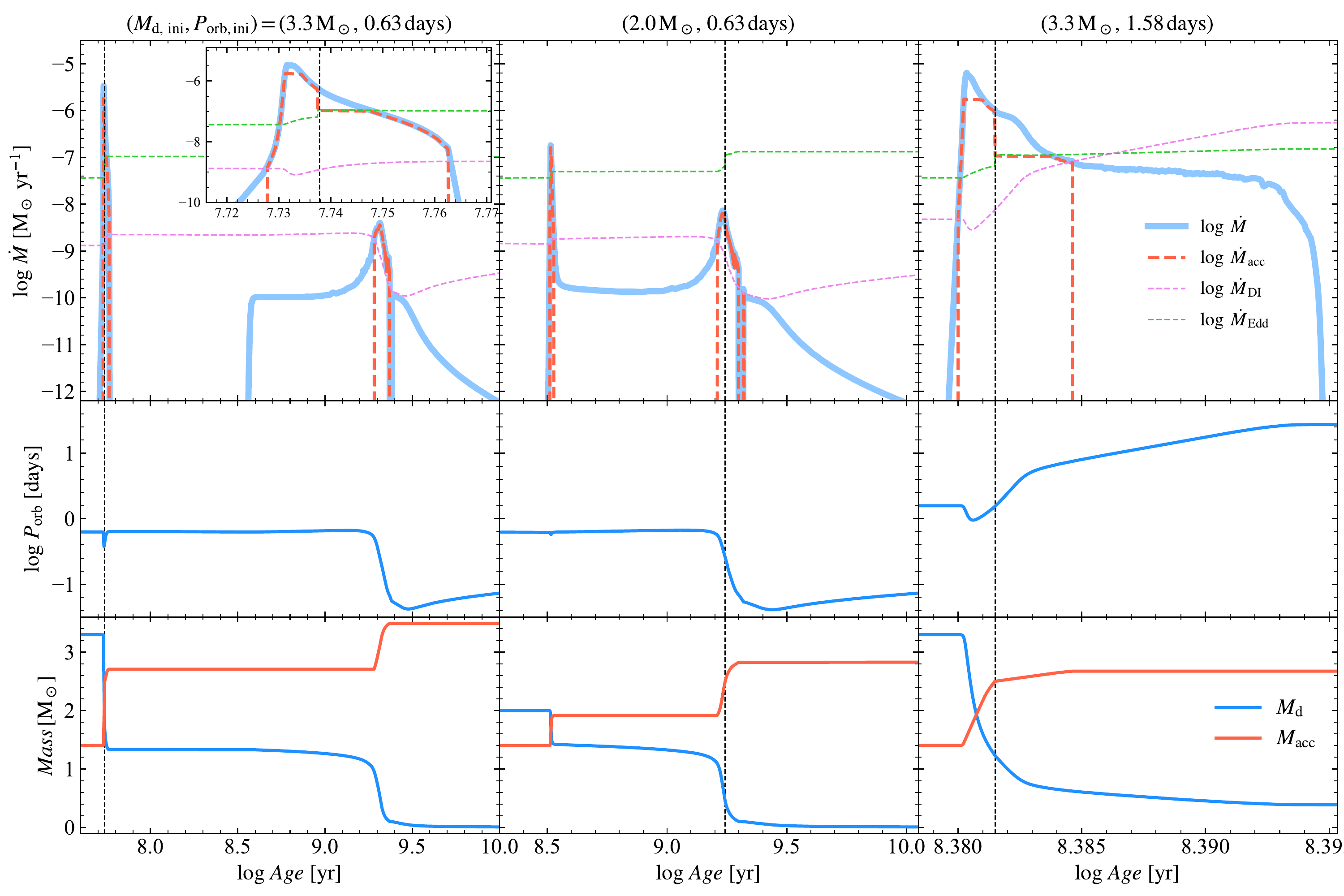}
	\caption{Three typical binary evolutionary tracks in Model~\textit{HB} with the initial parameters ($M_{\rm d,~ini}$, $P_{\rm orb,~ini}$)=($3.3~{\rm M_\odot}$, $0.63~{\rm day}$), ($2.0~{\rm M_\odot}$, $0.63~{\rm day}$) and ($3.3~{\rm M_\odot}$, $1.58~{\rm days}$), corresponding to the left, middle, and right panels, respectively. The upper, middle and lower panels depict the evolution of the mass transfer rate/mass accretion rate (blue solid lines/red dashed lines), the orbital period and the donor/accretor mass (blue/red solid lines) against the age, respectively. The purple and the green dashed lines in the upper panels represent the critical mass transfer rate of the DI and the Eddington accretion rate of the accretor, respectively. The black dashed lines indicate the ages when NSs become BHs.}
	\label{fig:tracks}
\end{figure*}

\quad It is traditionally thought that the accretion rate of a CO is limited by the Eddington limit, namely,
\begin{equation}\label{eq:Medd}
	\dot M_{\rm Edd}=3.64\times10^{-8}\left(\frac{1+X}{1.7}\right)^{-1}\left(\frac{M_{\rm acc}}{1.4~\rm M_\odot}\right)\left(\frac{\eta}{0.1}\right)^{-1}~{\rm M_{\odot}~yr^{-1}},
\end{equation}
where $M_{\rm acc}$ is the accretor's mass; $X$ is the hydrogen abundance of the accreting material; $\eta$ is the radiation efficiency to correct the accreted mass for the energy lost by radiation and we set $\eta=0.1$ for NSs and $\eta=1-\sqrt{1-({M_{\rm acc}}/{2.5~{\rm M_\odot}})^2/9}$ for BHs \citep{Podsiadlowski+2003}. For a super-Eddington BH accretion disk, \citet{Shakura+1973} suggested that the local Eddington-limited accretion is enabled and there is mass loss from the disk starting at the spherization radius
\begin{equation}
    R_{\rm sph}\simeq \left(\frac{27}{4}\right)\left(\frac{\dot M}{\dot M_{\rm Edd}}\right)\left(\frac{GM_{\rm acc}}{c^2}\right),
\end{equation}
where $\dot M$ is the mass accretion rate at the outer edge of the accretion disk (which is equal to the mass transfer rate $\dot M_{\rm L1}$ via RLOF), $c$ the speed of light in vacuum, and $G$ the gravitational constant.

The accretion behaviour of NSs is more complicated than that of BHs due to the interaction between the NS magnetic field and the accretion disk. When the NS magnetic field is sufficiently strong, the inner part of the disk is disrupted at a radius \citep{Frank+2002}
\begin{equation}\label{eq:R0}
  R_0=\xi \left(\frac{\mu^2}{\dot M_{\rm in}\sqrt{2GM_{\rm NS}}}\right)^{2/7},
\end{equation}
{where $\mu=BR_{\rm NS}^3$ is the magnetic moment of the NS (with $B$ and $R_{\rm NS}$ being the surface magnetic field and the radius respectively), $M_{\rm NS}$ the mass of the NS, $\dot M_{\rm in}$ the accretion rate at $R_0$, and $\xi$ a factor of order unity \citep{Ghosh+1979b,Wang+1996,Long+2005,Kulkarni+2007}. If $R_0>R_{\rm sph}$, then $\dot M_{\rm in}=\dot M$; otherwise $\dot M_{\rm in}=\dot M(R_0/R_{\rm sph})$. This limits $\dot M_{\rm in}$ to a critical accretion rate $\dot M_{\rm crit}$ by equating 
$R_0$ with $R_{\rm sph}$ \citep[e.g.,][]{Kuranov+2020}},
\begin{equation}\label{eq:mdotc}
	\begin{split}
	    \dot M_{\rm crit}&\simeq5.1\times10^{-7}\left(\frac{\xi}{0.5}\right)^{7/9}\left(\frac{\dot M_{\rm Edd}}{1.5\times10^{18}~{\rm g~s^{-1}}}\right)^{7/9}\\
	    &\times\left(\frac{\mu}{10^{30}~{\rm G~cm^3}}\right)^{4/9}\left(\frac{M_{\rm NS}}{1.4~{\rm M_\odot}}\right)^{-8/9}~{\rm M_\odot~{yr}^{-1}}.
	\end{split}
\end{equation}
The \citet{Shakura+1973} model is based on the assumption of a standard thin disk. More recently, \cite{Chashkina+2017} considered geometrically thick disk accretion onto a magnetized NS and pointed out that, in the case of $R_0>R_{\rm sph}$ the local Eddington limit is reached when the half-thickness of the disk is comparable with the inner radius of the disk. This gives
\begin{equation}
\dot M_{\rm crit,~1}\simeq 35~\dot M_{\rm Edd}\left(\frac{\mu}{10^{30}~{\rm G~cm^3}}\right)^{4/9}. 
\end{equation}
At a even higher accretion rate, advection becomes important \citep{Lipunova+1999} and the critical accretion rate can be enhanced to be \citep{Chashkina+2019}\footnote{Note that the magnitude of $\dot M_{\rm crit}$ also depends on other parameters including the viscosity parameter, the vertical effective polytropic index of the accretion disk, and the energy fraction for launching the disk winds.}
\begin{equation}
\dot M_{\rm crit,~2}\simeq 200~\dot M_{\rm Edd}\left(\frac{\mu}{10^{30}~{\rm G~cm^3}}\right)^{4/9}.
\end{equation}
Based on \cite{Chashkina+2017,Chashkina+2019} we assume that the upper limit of $\dot M_{\rm in}$ is
\begin{equation}\label{eq:mdotcch}
    \dot M_{\rm crit}\simeq
    \begin{cases}
        \dot M_{\rm crit,~1}
        \quad\quad{\rm if}~\dot M<\dot M_{\rm crit,~2},\\
        \dot M_{\rm crit,~2}
        \quad\quad{\rm if}~\dot M>\dot M_{\rm crit,~2}.
    \end{cases}
\end{equation}

We let the NS accretion rate $\dot M_{\rm NS}$ equal to $\dot M_{\rm in}$. As mentioned in last section, observations of ULX pulsars suggest that the NSs in ULXs may accrete at an accretion rate up to $\sim 10^2-10^3\dot M_{\rm Edd}$. Super-Eddington luminosities have also been reported in RLOF NS HMXBs like SMC X-1 \citep{Price+1971} and LMC X-4 \citep{Moon+2003}. This could be related to the strong magnetic fields of the NSs, which reduce the electron scattering cross section (\citealt{Canuto+1971,Herold+1979,Paczynski+1992}) and enhance the Eddington luminosity. Indeed, recent studies on the magnetic fields of ULX pulsars indicate that they may possess magnetic fields $\gtrsim 10^{11}-10^{15}~{\rm G}$ \citep{Dall+2015,XuKun+2017,Erkut+2020,Gao+2021}. In that case the accretion flow is collimated by the field lines, and radiation escapes from the side of the accretion column above the polar regions \citep[e.g.,][]{Basko+1976a,Basko+1976b,Mushtukov+2015}. Recent numerical simulations \citep[e.g.,][]{Kawashima+2016,Takahashi+2017,Kawashima+2020} also confirm that magnetized NSs can accrete at a rate much higher than the Eddington limit accretion rate.

We tentatively ignore possible magnetic field decay in our calculations, because it is still unclear whether it is caused by spindown-induced flux expulsion, Ohmic evolution of crustal field and diamagnetic screening of the field by accreted plasma
\citep[see][for a review]{Bhattacharya2002}. In the latter cases, we note that some LMXBs like 4U1626$-$67 possess an very old NS that has experienced extensive mass accretion, but the NS magnetic field is still as strong as a few $10^{12}~{\rm G}$ \citep{Verbunt+1990,Orlandini+1998,Camero+2012,Sasano+2014}. We will discuss the influence of magnetic field decay in Section~\ref{section:dis&sum}.

Another issue needed to address is the spin evolution of accreting NSs. Stable accretion requires that the inner disk radius $R_0$ should be less than the corotation radius 
\begin{equation}
    R_{\rm c}=\left(\frac{GM_{\rm NS}P_{\rm s}^2}{4\pi^2}\right)^{1/3},
\end{equation} 
where $P_{\rm s}$ is the spin period. When $R_0>R_{\rm c}$, the NS enters the propeller regime with most of the accreting material being ejected. This means that efficient accretion can take place if
\begin{equation}\label{eq:prop}
    \begin{split}
       \dot M_{\rm in}>\dot M_{\rm prop}&\simeq1.19\times10^{-9}\left(\frac{\mu}{10^{30}~{\rm G~cm^{3}}}\right)^2\\
       &\times\left(\frac{\xi}{0.5}\right)^{7/2}\left(\frac{M_{\rm NS}}{1.4~{\rm M_\odot}}\right)^{-5/3}\left(\frac{P_{\rm s}}{1~{\rm s}}\right)^{-7/3}~{\rm M_\odot~yr^{-1}}.
    \end{split}
\end{equation}
In our calculation, the NS mass growth generally satisfies this requirement. We can also check this condition by examining the spin evolution of the NSs. The timescale $\tau_{\rm s}$ of the spin evolution of accreting NSs can be roughly estimated from the torque equilibrium equation 
\begin{equation}
    2\pi I|\dot P_{\rm s}|/P^2_{\rm s}\simeq \dot M_{\rm in} (GM_{\rm NS}R_0)^{1/2},
\end{equation}
where $I$ is the moment of inertia and $\dot P_{\rm s}$ is the spin period derivation of the NS. So, it follows that
\begin{equation}\label{eq:tauspin}
	\begin{split}
		\tau_{\rm s}\sim\frac{P_{\rm s}}{|\dot P_{\rm s}|}
		&\sim52\left(\frac{\xi}{0.5}\right)^{-1/2}\left(\frac{M_{\rm NS}}{1.4~{\rm M_\odot}}\right)^{4/7}\left(\frac{R_{\rm NS}}{10^6~{\rm cm}}\right)^2\left(\frac{P_{\rm s}}{1~{\rm s}}\right)^{-1}\\
		&\times\left(\frac{\dot M}{10^{-6}~{\rm M_\odot~yr^{-1}}}\right)^{-6/7}\left(\frac{\mu}{10^{30}~{\rm G~cm^3}}\right)^{-2/7}~{\rm yr}.
	\end{split}
\end{equation}
The measured spin evolution timescales for the ULX pulsars M82~X-2 \citep{Bachetti+2014,Bachetti+2020}, NGC5907~ULX-1 \citep{Israel+2017+NGC5907ulx1}, M51~ULX-7 \citep{Rodrguez+2020} and NGC7793~P13 \citep{Furst+2016,Furst+2018,Israel+2017+NGC7793P13} are $\sim 200$, $50$, $40$ and $300~{\rm yr}$, respectively.
Since the mass transfer timescale $\tau_{\rm mt}\sim M_{\rm d}/\dot M\sim 10^5-10^{9}~{\rm yr}$ (where $M_{\rm d}$ is the donor's mass) is much longer than $\tau_{\rm s}$, we expect that the spin equilibrium can be quickly established once the mass transfer starts and the NS is always close to the spin equilibrium state during most of its X-ray lifetime \citep[e.g.,][]{Kuranov+2020}. The observed alternation between spin-up and spin-down of M82~X-2 is consistent with this picture \citep{Bachetti+2020}.

We assume the maximum mass of an NS to be $2.5~{\rm M_\odot}$, slightly higher than estimated by \cite{Farr+2020} for Galactic NSs. Once the NS accretes enough material and collapses into a BH\footnote{We neglect any kick caused by the collapse of an NS into a BH.}, the magnetic field effect stops working. As a result, the critical accretion rate recovers to the traditional Eddington limit, i.e., $\dot M_{\rm crit}=\dot M_{\rm Edd}$\footnote{As shown below, the descendant BHs mostly accrete at a sub-Eddington rate, so the definition of $\dot{M}_{\rm crit}$ does not considerably affect the properties of BHs.}.

A significant part of LMXBs are transient sources, likely caused by thermal and viscous instability in the accretion disks \citep{Lasota+2001,Hameury+2020R}. The condition for the disk instability (DI) is that the mass transfer rate in the disks is lower than a critical value $\dot M_{\rm DI}$, which depends on the masses of the accretor and the donor, the orbital period, and the chemical compositions of the disk material. Additionally, X-ray irradiations from the accreting NS/BHs can help stabilize the accretion disk to some extent by increasing the surface temperature of the disk \citep{Paradijs+1996,King+1997,Lasota+2001,Ritter+2008}. Here we use the prescription of $\dot M_{\rm DI}$ given by \cite{Dubus+1999} \citep[see also][]{Lasota+2008,Coriat+2012}, i.e.,
\begin{equation}\label{eq:DIM}
	\begin{split}
	\dot M_{\rm DI}\simeq
	&3.2\times10^{-9}\\
	&\times\left(\frac{M_{\rm acc}}{1.4~{\rm M_\odot}}\right)^{0.5}\left(\frac{M_{\rm d}}{1.0~{\rm M_\odot}}\right)^{-0.2}\left(\frac{P_{\rm orb}}{1~{\rm day}}\right)^{1.4}~{\rm M_\odot~yr^{-1}},
	\end{split}
\end{equation}
where $P_{\rm orb}$ is the binary orbital period. The accretion disk becomes unstable when $\dot M<\dot M_{\rm DI}$, that is, the disk undergoes rapid mass transfer during short outbursts separated by long quiescent intervals. \cite{Hameury+2020} calculated the time evolution of transient BH accretion disks, and found that the ratio of the accreted mass to the disk mass during outbursts is generally $\lesssim 0.05$. In the case of transient NS accretion disks, it is usually thought that the disk is passive during quiescence and the NSs are in the propeller or ejector regime, so almost all of the transferred material is likely to be ejected. During outbursts, the mass flow rate is enhanced by a factor of $\sim 10^2-10^3$ \citep{Tanaka+1996,King+2003}, and some of the disk material may be accreted by the NS. However, the accretion efficiency is sensitively dependent on the properties of the NS and the accretion disk. Fortunately, since we are only concerned with DI at the very beginning of the mass transfer of IMXBs\footnote{NSs in transient LMXBs can hardly evolve to be BHs due to the low mass of the donor stars.} and the initial rising stage occupies a small part in the total mass transfer lifetime (see \autoref{fig:tracks}), the contribution of mass accretion during outbursts to the NS mass growth is limited. 
For simplicity, we assume that mass accretion of the CO is fairly inefficient from an unstable disk, and set the ratio of the lost mass to the transferred mass in unit time as follows,
\begin{equation}\label{eq:beta}
	\beta_{\rm mt}=
	\begin{cases}
		{(\dot M-\dot M_{\rm crit})}/{\dot M},&{\rm if}~{\dot M>\dot M_{\rm crit}~\&~\dot M>\dot M_{\rm DI}}\\
		{0,}&{\rm if}~{\dot M_{\rm DI}<\dot M<{\dot M_{\rm crit}}}\\
		{1,}&{\rm if}~{\dot M<\dot M_{\rm DI}}
	\end{cases}.
\end{equation}
Hence, the accretion rate of the CO is $\dot M_{\rm acc}=(1-\eta)(1-\beta_{\rm mt})\dot M$. (Note that $\dot M_{\rm crit}$ in \autoref{eq:beta} is different for NSs and BHs.)
In Section~\ref{section:dis&sum} we will check this assumption by taking $\beta_{\rm mt}=0.5$ for NSs.

Mass loss also takes away orbital angular momentum from the binary system \citep{Soberman+1997}. We assume that mss loss occurs in the vicinity of the accretor in I/LMXBs, so the rate of angular momentum loss can be written as
\begin{equation}
	\dot J_{\rm ml}=-\beta_{\rm mt}\dot M \left(\frac{M_{\rm acc}a}{M_{\rm d}+M_{\rm acc}}\right)^2\frac{2\pi }{P_{\rm orb}},
\end{equation}
where $a$ is the orbital separation. 
{Our assumption of non-conservative mass transfer in transient LMXBs obviously influence the angular momentum loss from the binary compared with the case of conservative mass transfer. However, as shown by \cite{2009ApJ...691.1611M}, the strength of magnetic braking (see below) is the dominant factor in determining the value of the bifurcation periods compared with mass loss. Thus, the amount and form of the ejected mass does not considerably affect the evolution of BH LMXBs.}

For wind mass loss from the low-mass donor, we consider angular momentum loss caused by braking of the magnetized wind. We use the convection-boosted magnetic braking model suggested by \cite{Van+2019}, i.e.,
\begin{equation}
	\dot J_{\rm mb}=-3.8\times10^{-30}M_{\rm d}{\rm R^4_\odot}\left(\frac{R_{\rm d}}{\rm R_\odot}\right)^{\gamma_{\rm mb}}\Omega^3\left(\frac{\tau_{\rm c}}{\tau_{\rm \odot}}\right)^{\xi_{\rm mb}}~{\rm dyne~cm},
\end{equation}
where $\gamma_{\rm mb}\sim 0-4$ is the magnetic braking parameter (in our calculations we take the default value $\gamma_{\rm mb}=3$); $\Omega$ is the orbital angular velocity, $R_{\rm d}$ is the radius of the donor star; $\tau_{\rm c}$ is the turnover time of the convective eddies of the donor star and $\tau_{\rm \odot}\simeq2.8\times10^6~{\rm s}$ is the turnover time of the Sun; $\xi_{\rm mb}$ is the convection-boosted power low index and we take $\xi_{\rm mb}=2$ (\citealt{Van+2019,Deng+2021}). The convection-boosted magnetic braking model is based on the widely-used traditional magnetic braking model \citep{Rappaport+1983} by adding a boosting factor $\tau_{\rm c}$. \cite{Van+2019} found that this model is able to reproduce some particular NS LMXBs. \cite{Deng+2021} systematically compared five types of magnetic braking prescriptions in the literature, and found that the convection-boosted model seems to be more suitable to account for the overall properties of LMXBs, compared with other magnetic braking prescriptions. 

Finally we also take into account angular momentum loss caused by GW emissions \citep{Landau+1959,Faulkner+1971}, i.e.,
\begin{equation}
	\dot J_{\rm gr}=-\frac{32}{5c^2}\left(\frac{2\pi G}{P_{\rm orb}}\right)^{7/3}\frac{M_{\rm d}M_{\rm acc}}{(M_{\rm d}+M_{\rm acc})^{2/3}}.
\end{equation}

We then use the one-dimension stellar evolution code \textsc{MESA} \citep[version 12778,][]{MESA+2011,MESA+2013,MESA+2015,MESA+2018,MESA+2019} to trace the detailed evolutions of I/LMXBs. For the incipient I/LMXBs, we consider the initial mass of the NS to be $1.4~{\rm M_\odot}$, and assume that the donor evolves from zero-age main sequence, since the lifetime ($\lesssim 10^7~{\rm yr}$) of the NS progenitor is much shorter than that of a low-mass star. The initial mass of the donor is distributed over the range of $0.4-6~{\rm M_\odot}$ in linear steps of $0.1$, and the orbital period over the range of $10^{-1}-10^4~{\rm days}$ in logarithmic steps of $0.1$, consistent with the matrix elements in \autoref{fig:birthrate2d}. We terminate the calculation when the age of the donor star reaches $12~{\rm Gyr}$. We exclude those binaries without mass transfer, with RLOF at the birth of the NS, and with the donor star overflowing from the $L_2$ Lagrange point.

We use a simple criteria, that is, whether the mass transfer rate $\dot M$ is less than $10^{-4}~{\rm M_\odot~yr^{-1}}$ or not, to discriminate the dynamically stable or unstable mass transfer in I/LMXBs. To explore the possible influence of the magnetic field, metallicity, and critical mass transfer rate on the mass growth of NSs, we construct the following six models:
\begin{itemize}
    \item Model \textit{LB}:
    $B=1\times10^{12}~{\rm G}$, $Z=0.02$ (case~1), and $\dot M_{\rm crit}$ follows \autoref{eq:mdotc} ($\simeq 20\dot M_{\rm Edd}$).
    \item Model~\textit{HB}:
    $B=1\times10^{13}~{\rm G}$, $Z=0.02$ (case~1), and $\dot M_{\rm crit}$ follows \autoref{eq:mdotc} ($\simeq 50\dot M_{\rm Edd}$).
    \item Model~\textit{LZ}:
    $B=1\times10^{12}~{\rm G}$, $Z=0.001$ (case~2), and $\dot M_{\rm crit}$ follows \autoref{eq:mdotc} ($\simeq 20\dot M_{\rm Edd}$).
    \item Model~\textit{LBCh}:
    $B=1\times10^{12}~{\rm G}$, $Z=0.02$ (case~1), and $\dot M_{\rm crit}$ follows \autoref{eq:mdotcch}.
    \item Model~\textit{HBCh}:
    $B=1\times10^{13}~{\rm G}$, $Z=0.02$ (case~1), and $\dot M_{\rm crit}$ follows \autoref{eq:mdotcch}.
    \item Model~\textit{LZCh}:
    $B=1\times10^{12}~{\rm G}$, $Z=0.001$ (case~2), and $\dot M_{\rm crit}$ follows \autoref{eq:mdotcch}.
\end{itemize}
The input parameters of the six models are summarized in \autoref{tab:models}.
\begin{table}
	\centering
	\setlength{\tabcolsep}{12pt}
	\caption{Different models in our calculation.\label{tab:models}}
	\begin{tabular}{cccc}
		\hline
        Models &$B~[{\rm G}]$&$Z$&$\dot M_{\rm crit}$\\
		\hline
        LB &$1\times10^{12}$&$0.02$  (case~1)&\autoref{eq:mdotc}\\
        HB &$1\times10^{13}$&$0.02$ (case~1)&\autoref{eq:mdotc}\\
        LZ &$1\times10^{12}$&$0.001$ (case~2)&\autoref{eq:mdotc}\\
        LBCh &$1\times10^{12}$&$0.02$ (case~1)&\autoref{eq:mdotcch}\\
        HBCh &$1\times10^{13}$&$0.02$ (case~1)&\autoref{eq:mdotcch}\\
        LZCh &$1\times10^{12}$&$0.001$ (case~2)&\autoref{eq:mdotcch}\\
		\hline
	\end{tabular}
\end{table}

\autoref{fig:stablemt} shows the allowed parameter space for stable mass transfer in different models. The colors in each pixel indicate the birthrates of specific I/LMXBs inferred from the BPS results. We use the blue and red dashed contours in each panel of \autoref{fig:stablemt} to indicate the final masses of the accretors $M_{\rm acc,~f}=2.5~{\rm M_\odot}$ and $3.0~{\rm M_\odot}$, respectively, and the green dashed contours to indicate $M_{\rm acc,~f}=2.5~{\rm M_\odot}$ with the Eddington-limited accretion prescription for comparison. Since the critical transfer rate increase with increasing magnetic field, NSs with a higher magnetic field have more opportunities to collapse into BHs, so the parameter space for the formation of mass-gap BHs enlarges with increasing magnetic field. It is also clear that mass-gap BHs can be more likely formed in the thick disk model (\autoref{eq:mdotcch}) than in the thin disk model (\autoref{eq:mdotc}) for the critical mass transfer rate. Our results in case~1 are generally in accordance with \cite{Tauris+2000} and \cite{Shao+2012MSP,Shao+2015}. This means that allowing supper-Eddington accretion does not strongly influence the conditions for stable mass transfer in I/LMXBs.

Lower metallicity leads to smaller radius of the donor star with a given mass. As a result, the maximum orbital periods with which RLOF can occur in Model~\textit{LZ} are shorter than in Model~\textit{LB}. This explains why the maximum mass of the donor in Model~\textit{LZ} is smaller than in Model~\textit{LB} for stable mass transfer. The black dotted lines in \autoref{fig:stablemt} represent the bifurcation period $P_{\rm bif}$ which divides diverging binaries from contracting binaries \citep{Pylyser+1988}, as a function of the initial donor mass $M_{\rm d,~ini}$. Note that the reduction of the metallicity results in a smaller radius of stars and shorter nuclear timescales which leads to a shorter $P_{\rm bif}$ \citep[see e.g.,][]{Jia+2014}. We find that an NS in Models~\textit{LB} and \textit{HM} can grow to be a BH if the initial orbital periods $\lesssim 10~{\rm days}$ and the initial donor masses $\gtrsim1.7~{\rm M_\odot}$, while in Model~\textit{LZ} the formation of a BH requires a higher initial donor mass. BHs with masses $>3~{\rm M_\odot}$ mostly originate from converging binaries in case~1 with the initial donor mass $\gtrsim 2~{\rm M_\odot}$. No BH with mass $>3~{\rm M_\odot}$ can form in case~2.

\autoref{fig:tracks} shows three typical binary evolutionary tracks in Model~\textit{HB} with the initial parameters ($M_{\rm d,~ini}$, $P_{\rm orb,~ini})=(3.3~{\rm M_\odot}$, $0.63~{\rm day}$), ($2.0~{\rm M_\odot}$, $0.63~{\rm day}$) and ($3.3~{\rm M_\odot}$, $1.58~{\rm days}$), corresponding to the left (track~1), middle (track~2) and right (track~3) panels, respectively. The initial states of the three tracks are shown in the upper middle panel of \autoref{fig:stablemt}. The upper, middle and lower panels depict the evolution of the mass transfer rate/mass accretion rate, the orbital period and the donor/accretor mass against the age, respectively. The left panels of \autoref{fig:tracks} demonstrate the evolution of a converging binary. The $3.3~{\rm M_\odot}$ donor starts to fill its RL and initiates mass transfer at the age of $\sim 51.37~{\rm Myr}$. The mass transfer rate goes up rapidly and exceeds the Eddington limit at $\sim53.71~{\rm Myr}$. At $\sim 54.68~{\rm Myr}$, the NS collapses to be a BH. At this moment, the donor mass is $\sim 1.69~{\rm M_\odot}$ and the orbital period is $\sim 0.44~{\rm day}$. After the formation of the BH, the orbit slightly expands due to the mass redistribution between the donor and the accretor, and the binary becomes detached. When the donor re-fills its RL at $\sim 0.37~{\rm Gyr}$, a BH XRB emerges. The subsequent mass transfer rate varies in the range of $\sim 10^{-10}-10^{-8}~{\rm M_\odot~yr^{-1}}$. Ultimately, the donor star evolves to be a brown dwarf (with mass $\sim 0.008~{\rm M_\odot}$), accompanied by a $\sim 3.49~{\rm M_\odot}$ BH in a $1.85~{\rm hour}$ orbit. If we take the Eddington-limited accretion prescription for the NS instead, the final NS mass in the binary can only reach $\sim 2.4~{\rm M_\odot}$.

In the middle panels of \autoref{fig:tracks}, the $2~{\rm M_\odot}$ donor overflows its RL at $\sim 319.64~{\rm Myr}$. The mass accretion rate of the NS varies in the range of $\sim 10^{-9}-10^{-7}~{\rm M_\odot~yr^{-1}}$ and the mass accretion process lasts about $333~{\rm Myr}$. When the NS mass increases to $\sim 2~{\rm M_\odot}$, the mass growth ceases because the disk becomes unstable. Subsequently, magnetic braking drives the mass transfer at an increasing rate and the NS begins to accrete material again. At $1.75~{\rm Gyr}$, the NS collapses to be a BH. At $\sim2~{\rm Gyr}$, the donor becomes fully convective and magnetic braking switches off, so the donor detaches from its RL. About $90.91~{\rm Myr}$ later, the donor fills its RL again driven by GW emissions, but the BH mass hardly increases. In the end, the binary consists of a $2.83~{\rm M_\odot}$ BH and a $0.0078~{\rm M_\odot}$ brown dwarf in a $1.79~{\rm hour}$ orbit.

The right panels of \autoref{fig:tracks} reflect the evolution of a diverging binary. The donor fills its RL at $\sim 239.7~{\rm Myr}$. After less than $1.01$-${\rm Myr}$ rapid accretion, the NS collapses into a BH. After the reversion of the component masses, the orbit period continually increases caused by the nuclear expansion of the donor star. The final orbital period is $27.36~{\rm days}$, and the donor ultimately becomes a $\sim 0.39~{\rm M_\odot}$ hybrid (Helium and Carbon) WD. However, the NS hardly accretes any material ($\lesssim 0.044~{\rm M_\odot}$) with the Eddington-limited accretion prescription.

\begin{figure*}
	\centering
	\includegraphics[width=\linewidth]{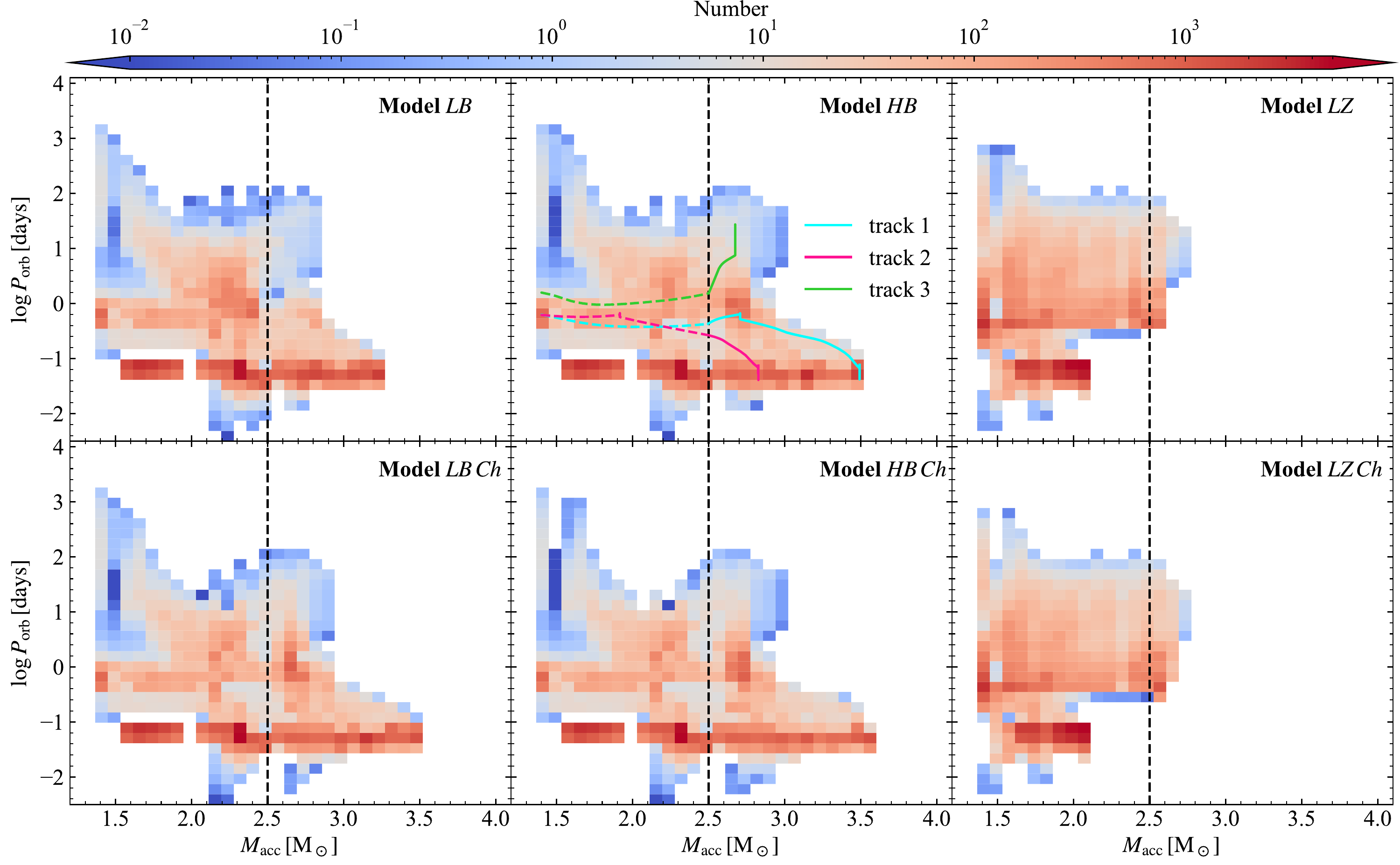}
	\caption{The number distributions of I/LMXBs in the accretor mass-orbital period ($M_{\rm acc}$-$P_{\rm orb}$) plane in different models. The color in each pixel indicates the number of I/LMXBs. The dashed lines ($M_{\rm acc}=2.5~{\rm M_\odot}$) show the dividing line between NS and BH accretors. The cyan, pink and green solid (dashed) lines in the upper middle panel show the evolutionary tracks~1, 2 and 3 in the $M_{\rm acc}$-$P_{\rm orb}$ plane, respectively, when they are in the BH (NS) XRB phase.}
	\label{fig:MaccPorb}
\end{figure*}

\section{Population results of mass-gap BH binaries}\label{section:results}

\quad Combining the detailed evolutionary sequences and the birthrate from the BPS, we can calculate the number distribution of I/LMXBs in $(A,~B)$ parameter plane. The number in a parameter space bin $(A_i,~B_j)$ can be obtained with the following equation
\begin{equation}
	N_{i,j}=\sum^{N_{\rm MESA}}_{k=1} T_{i,j,k}\mathcal{R}_{i,j,k},
\end{equation}
where $T_{i,j,k}$ and $\mathcal{R}_{i,j,k}$ are the time span and the birthrate for the $k$th evolutionary track passing through a specific parameter space element $(A_i,~B_j)$, respectively; $N_{\rm MESA}$ is the total number of the evolutionary tracks calculated with \textsc{MESA} that pass through the specific element $(A_i,~B_j)$. Here $A_i$ and $B_i$ can be any parameters we are interested in, such as the donor mass $M_{\rm d}$, the orbital period $P_{\rm orb}$, and the mass transfer rate $\dot M$. Based on this method, one can derive the number distribution of binaries in different evolutionary phases and in different parameter space.

We plot the number distributions of I/LMXBs in the accretor mass-orbital period ($M_{\rm acc}$-$P_{\rm orb}$) plane in \autoref{fig:MaccPorb}. The cyan, pink and green solid (dashed) lines in the upper left panel show the evolutionary tracks~1, 2 and 3 in the $M_{\rm acc}$-$P_{\rm orb}$ plane, respectively, when they are in the BH(NS) XRB phase. The black dashed line in \autoref{fig:MaccPorb} represents the dividing line between the NS and BH accretors. We can see that the accretor masses are distributed in the range of $\sim 1.4-3.6~{\rm M_\odot}$, and a considerable fraction of them are regarded to be mass-gap BHs. The maximum mass that the accretor can reach increases with increasing magnetic field. We list the predicted birthrate and the total number of NS/BH XRBs in Tables~\ref{tab:br} and \ref{tab:number}, respectively. In case~1, most of the BH binaries evolve from converging systems. In case~2 (Models~\textit{LZ{ \rm and} LZCh}), both the maximum accretor mass and the number of the BH XRBs are significantly smaller than in case~1.

\begin{table}
	\centering
	\setlength{\tabcolsep}{10pt}
	\caption{The predicted birthrates of incipient NS I/LMXBs and the final NS and BH binaries in units of $\rm yr^{-1}$.\label{tab:br}}
	\begin{tabular}{cccc}
		\hline
		Models & NS I/LMXBs & NS binaries & BH binaries\\
		\hline
		LB &$4.05\times10^{-5}$&$3.38\times10^{-5}$&$6.69\times10^{-6}$\\
		HB &$4.28\times10^{-5}$&$2.89\times10^{-5}$&$1.39\times10^{-5}$\\
		LZ &$8.44\times10^{-5}$&$7.69\times10^{-5}$&$7.47\times10^{-6}$\\
		LBCh &$4.83\times10^{-5}$&$2.15\times10^{-5}$&$2.67\times10^{-5}$\\
		HBCh &$4.93\times10^{-5}$&$1.75\times10^{-5}$&$3.18\times10^{-5}$\\
		LZCh &$8.69\times10^{-5}$&$6.61\times10^{-5}$&$2.08\times10^{-5}$\\
		\hline
	\end{tabular}
\end{table}

\begin{table*}
	\centering
	\setlength{\tabcolsep}{8pt}
	\caption{Predicted numbers of NS/BH XRBs, ULXs, detached BH binaries and LISA BH binaries (with $SN>5$) in the Galaxy.\label{tab:number}}
	\begin{tabular}{cccccccc}
		\hline
		Models & Total XRBs & NS XRBs & NS ULXs & BH XRBs & BH ULXs & Detached BH binaries & LISA BH binaries ($SN>5$)\\
		\hline
		LB  &$6.29\times10^4$&  $4.34\times10^4$&   $32$&   $1.95\times10^4$&   $0$&    $9$&    $1$\\
		HB  &$6.28\times10^4$&  $3.99\times10^4$&   $27$&   $2.30\times10^4$&   $1$&    $47$&   $1$\\
		LZ  &$6.29\times10^4$&  $6.04\times10^4$&   $47$&   $2.50\times10^3$&   $0$&    $35$&   $0$\\
		LBCh    &$6.41\times10^4$&  $4.00\times10^4$&   $26$&    $2.41\times10^4$&   $4$&    $74$&   $2$\\
		HBCh    &$6.17\times10^4$&  $3.99\times10^4$&   $19$&    $2.29\times10^4$&   $8$&    $30$&   $3$\\
		LZCh    &$6.29\times10^4$&  $5.98\times10^4$&   $45$&    $3.12\times10^3$&   $1$&    $120$&  $0$\\
		\hline
	\end{tabular}
\end{table*}

In the following, we demonstrate the statistical characteristics of these BH binaries and their possible contribution to the GW sources. We may put useful constraints on the mass accretion efficiency in I/LMXBs by comparing their properties with multi-messenger observations.

\subsection{Orbital periods, donor masses, and mass transfer rates of BH I/LMXBs}

\begin{figure*}
	\centering
	\includegraphics[width=\linewidth]{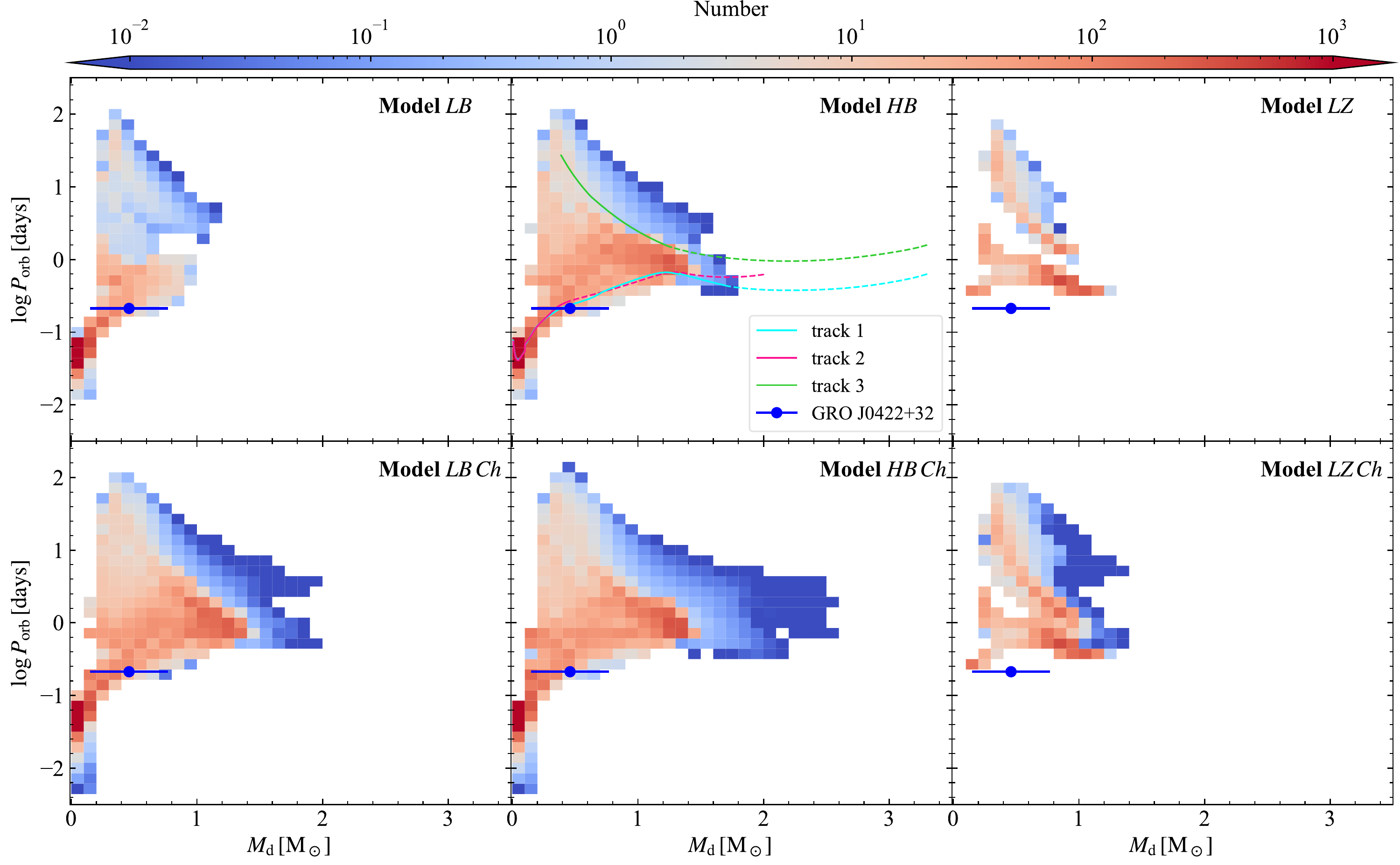}
	\caption{The number distribution of BH I/LMXBs in the donor mass-orbital period ($M_{\rm d}$-$P_{\rm orb}$) plane. The color in each pixel represents the number of BH I/LMXBs. The blue dot with error bar in each panel shows the position of GRO~J0422+32. The model parameters in each panel are the same as in \autoref{fig:MaccPorb}.}
	\label{fig:MdorPorbBH}
\end{figure*}

\quad In \autoref{fig:MdorPorbBH}, we show the number distribution of BH I/LMXBs in the donor mass-orbital period ($M_{\rm d}$-$P_{\rm orb}$) plane. The orbital periods of BH I/LMXBs are distributed in the range of $\sim 0.01-100~{\rm days}$ and there are some ultracompact binaries with very short orbital periods $\sim 0.01-0.1~{\rm days}$ and low-mass ($<0.1~{\rm M_\odot}$) donors. These binaries originate from the track~1/2 evolution and the donors are degenerated hydrogen/helium dwarf stars. They will evolve into the GW sources to be probably detected by future space-based GW detectors. The blue dot with error bar denotes the mass-gap BH LMXB GRO~J0422+32 (a $\gtrsim 2.1~{\rm M_\odot}$ BH accompanied by a $0.46\pm0.31~{\rm M_\odot}$ star in a $0.212~{\rm day}$ orbit, \citealt{Webb+2000,Gelino+2003,Kreidberg+2012}). Our results of case~1 evolution can properly match the donor mass and orbital period of GRO~J0422+32.

Metallicity can significantly influence the total number and number distribution of BH I/LMXBs. From \autoref{tab:number} we find that, although the birthrate of I/LMXBs in case~1 is about half of that in case~2, the total number of BH I/LMXBs in case~1 is an order of magnitude more than that in case~2. This difference results from the lower bifurcation period in case~2, which helps produce more converging systems in case~1. For comparison, Figure 5 shows that there are few BH I/LMXBs with orbital periods $\lesssim 0.5~{\rm day}$ in case~2. Hence, the DI is less likely to occur in case~1, so the NS can accrete more material than in case~2.

\begin{figure*}
	\centering
	\includegraphics[width=\linewidth]{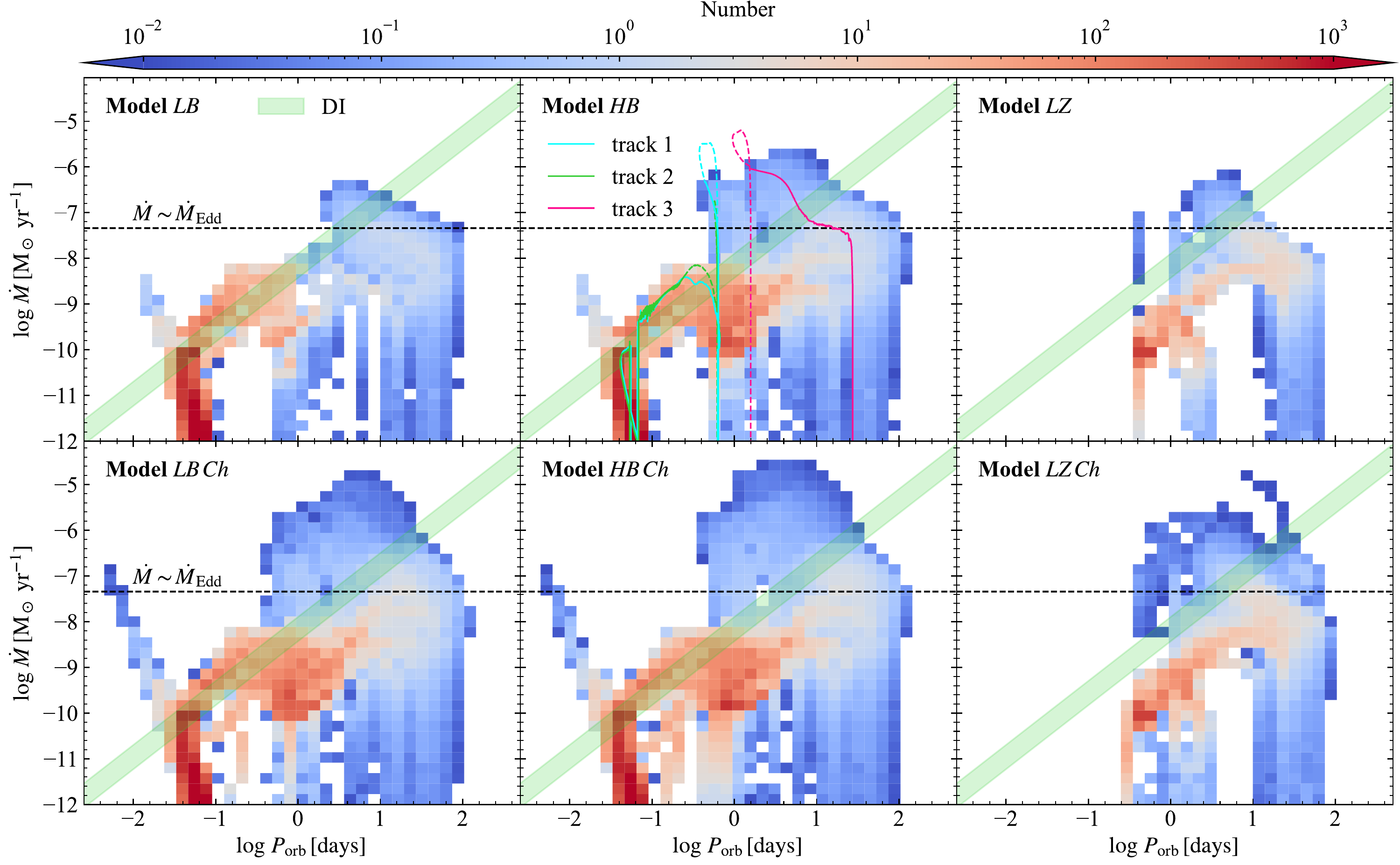}
	\caption{The number distribution of BH I/LMXBs in the orbital period-mass transfer rate ($P_{\rm orb}$-$\dot M$) plane. The green belt in each panel represents the dividing line (\autoref{eq:DIM}) between the persistent and transient XRBs. The model parameters in each panel are the same as in \autoref{fig:MaccPorb}.
	\label{fig:PorbMdotBH}}
\end{figure*}

\autoref{fig:PorbMdotBH} shows the number distribution of BH I/LMXBs in the orbital period-mass transfer rate ($P_{\rm orb}$-$\dot M$) plane. Combining the number distribution of BH I/LMXBs and the three evolutionary tracks in the upper middle panel of \autoref{fig:PorbMdotBH}, we find that the mass transfer in BH I/LMXBs maintains at a rate $\sim 10^{-10}-10^{-8}~{\rm M_\odot~yr^{-1}}$, which is dominated by the track~2 evolution and the late track~1 evolution. BH I/LXMBs with a relatively high mass transfer rate ($\gtrsim 10^{-8}~{\rm M_\odot~yr^{-1}}$) are usually in a relatively wide orbit ($P_{\rm orb}\gtrsim 1~{\rm day}$), mainly contributed by the early track~1 and track~3 evolutions. BH I/LMXBs with orbital period $\lesssim 0.1~{\rm day}$ and mass transfer rate $\lesssim10^{-10}~{\rm M_\odot~yr^{-1}}$ experience the later track~1/2 evolution driven by GW radiation. The green belt in each panel of \autoref{fig:PorbMdotBH} indicates the dividing line between the transient and persistent XRBs. We find that most of the mass-gap BH XRBs are expected to be transient X-ray sources.

\subsection{Luminosities of I/LMXBs and their contribution to ULXs}

\quad We use the following equation to calculate the isotropic accretion luminosities $L_{\rm acc}$ of I/LMXBs (\citealt{Shakura+1973}, see also \citealt{King+2017,King+2020}), i.e.,
\begin{equation}\label{eq:Lacc}
	L_{\rm acc}=\begin{cases}
		{\eta\dot M c^2,}&{\dot M<\dot M_{\rm crit}}\\
		{\eta\dot M_{\rm crit}c^2[1+\ln(\dot M/\dot M_{\rm crit})],}&{\dot M>\dot M_{\rm crit}}
	\end{cases}.
\end{equation}
Note that $\dot M_{\rm crit}$ is different in the cases of accreting NSs and BHs.

\begin{figure*}
	\centering
	\includegraphics[width=0.92\linewidth]{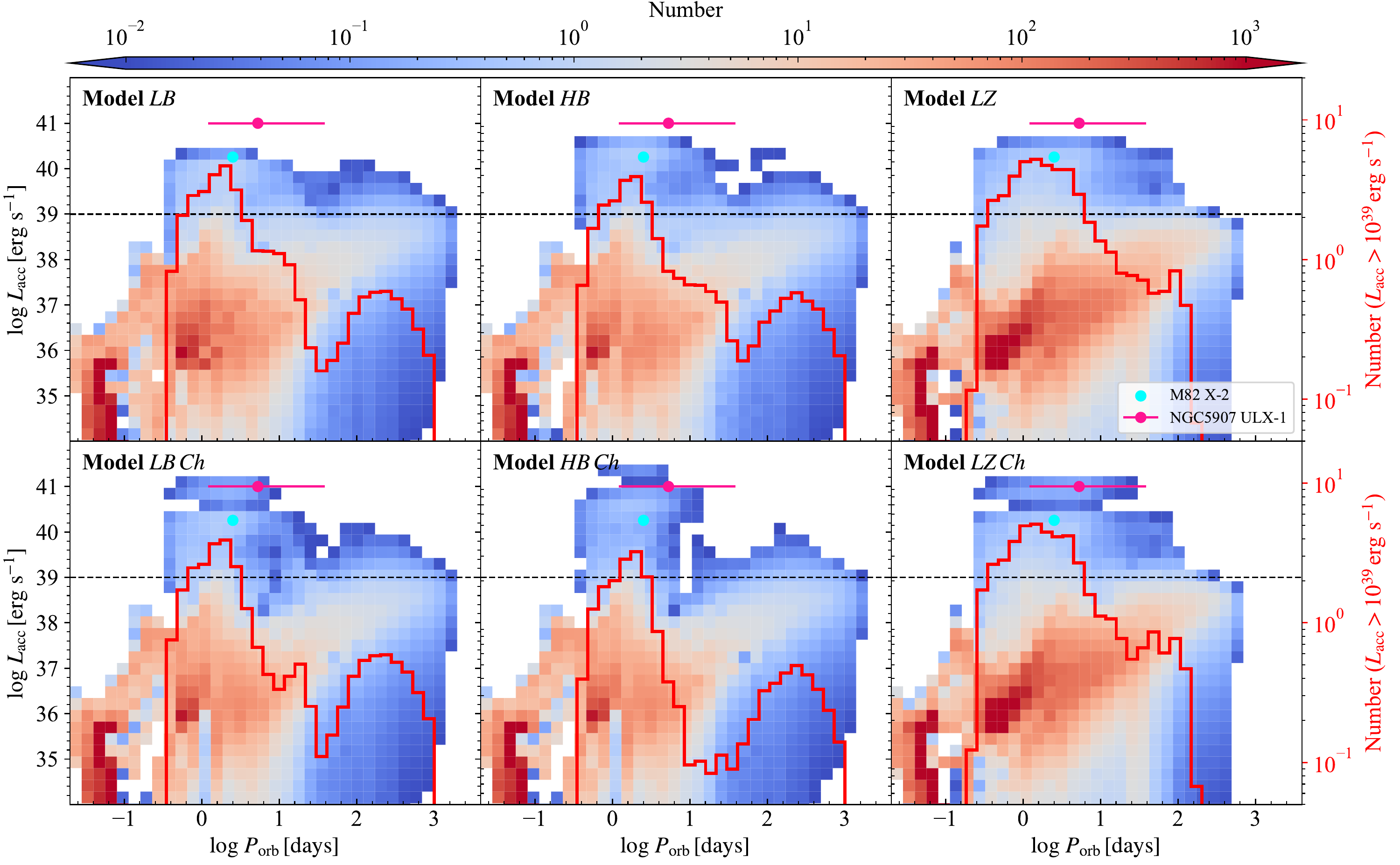}
	\caption{The number distribution of NS I/LMXBs in the orbital period-accretion luminosity ($P_{\rm orb}$-$L_{\rm acc}$) plane in the six models. The dashed line in each panel represents the criterion of ULXs as $L_{\rm acc}=10^{39}~{\rm erg~s^{-1}}$. The red histogram in each panel indicates the number distribution of NS ULXs as a function of the orbital period. The red and cyan circles show the observed X-ray luminosities and orbital periods of M82~X-2 and NGC5907~ULX-1, respectively. The model parameters in each panel are the same as in \autoref{fig:MaccPorb}.	\label{fig:PorbLaccNS}}
\end{figure*}

\begin{figure*}
	\centering
	\includegraphics[width=0.92\linewidth]{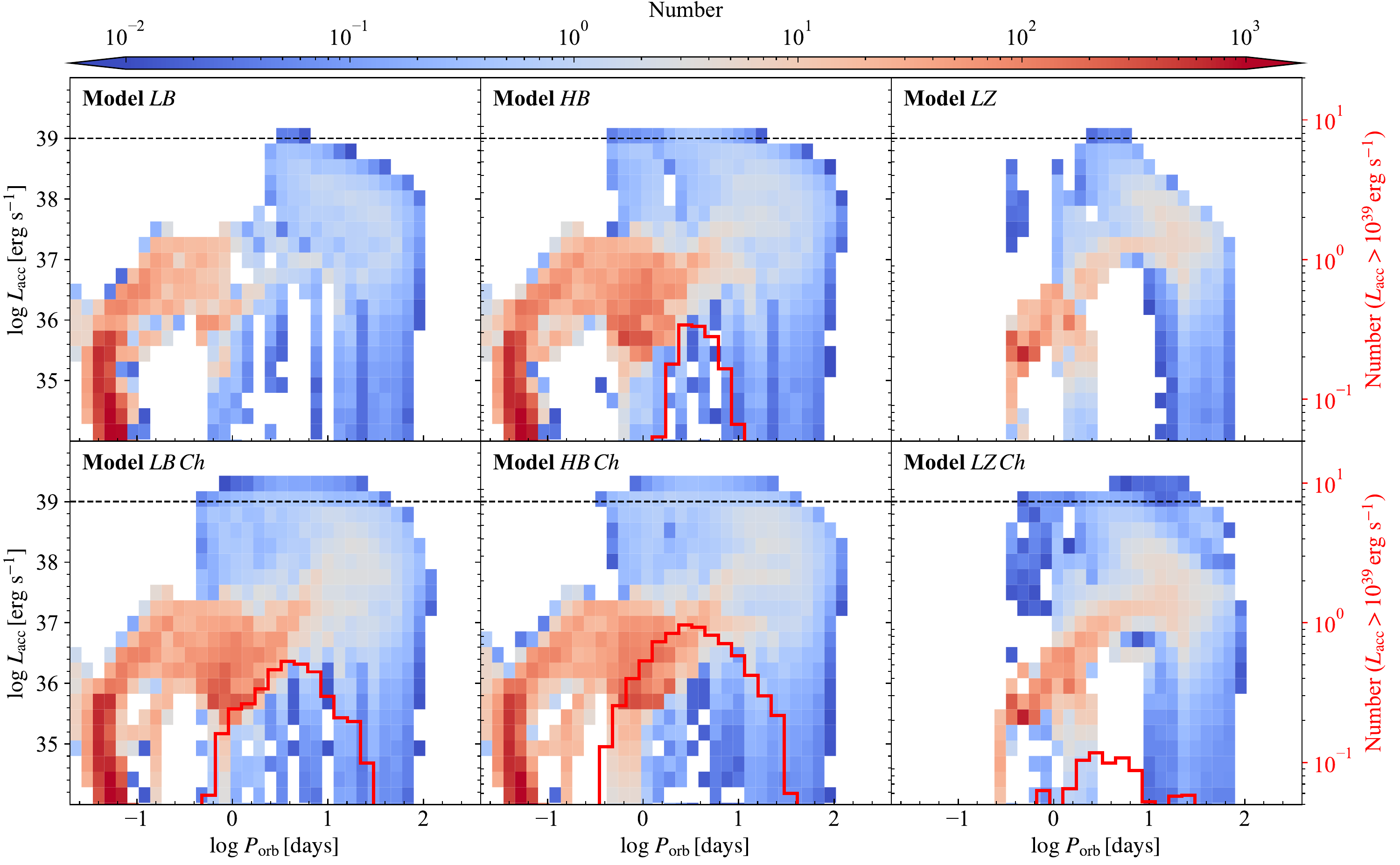}
	\caption{ Same as \autoref{fig:PorbLaccNS} but for BH I/LMXBs.\label{fig:PorbLaccBH}}
\end{figure*}

We plot the number distributions of NS and BH I/LMXBs in the orbital period-accretion luminosity ($P_{\rm orb}$-$L_{\rm acc}$) plane in Figures~\ref{fig:PorbLaccNS} and~\ref{fig:PorbLaccBH}, respectively. The luminosities of NS I/LMXBs are mostly in the range of $\sim 10^{34}-10^{39}~{\rm erg~s^{-1}}$ and a fraction of them can reach $\gtrsim 10^{40}~{\rm erg~s^{-1}}$, significantly exceeding the Eddington luminosity for an NS accretor. The luminosities of BH I/LMXBs are in the range of $\sim 10^{34}-10^{39}~{\rm erg~s^{-1}}$. Metallicity does not significantly influence the number distribution of NS I/LMXBs in the $P_{\rm orb}$-$L_{\rm acc}$ plane, but the total number of NS/BH I/LMXBs heavily depends on metallicity. From \autoref{tab:number}, we find that the number of NS I/LMXBs in case~2 is greater than that in case~1, because the birthrate of incipient NS I/LMXBs in case~2 is twice that in case~1.

Assuming that I/LMXBs with isotropic luminosities $L_{\rm acc}>10^{39}~{\rm erg~s^{-1}}$ (depicted with the dashed line in each panel of Figures~\ref{fig:PorbLaccNS} and~\ref{fig:PorbLaccBH}) are ULXs, we plot the number distribution of NS and BH ULXs as a function of the orbital period with red histograms in Figures~\ref{fig:PorbLaccNS} and~\ref{fig:PorbLaccBH}, respectively. In \autoref{tab:number}, we also list the number of NS and BH ULXs. In our results, there are $\sim 19-47$ NS ULXs and only $\sim 0-8$ BH ULXs in the Galaxy. The orbital periods of the NS and BH ULXs are in the range of $\sim 0.2-200~{\rm days}$ and $\sim5-50~{\rm days}$, respectively. The majority of NS ULXs originate from the early track~1/3 evolution in a relatively close orbit. They will finally evolve into BH XRBs. Other NS ULXs have relatively low-mass ($\lesssim 1~{\rm M_\odot}$) donors and long ($>10$ days) orbital periods, and it is hard for them to accrete enough material to collapse to be BHs.

M82~X-2 is a ULX pulsar with a $\sim 5.2~{\rm M_\odot}$ donor in a $2.53~{\rm days}$ orbit \citep{Bachetti+2014}. The extremely high luminosity (up to $\sim 10^{40}~{\rm erg~s^{-1}}$) implies that the NS in M82~X-2 is undergoing rapid mass accretion. NGC5907~ULX-1 \citep{Israel+2017+NGC5907ulx1} is another pulsar accompanied by a $\sim 2-6~{\rm M_\odot}$ donor in a $\sim 5.3~{\rm day}$ orbit. Its peak luminosity can reach $\sim 10^{41}~{\rm ergs^{-1}}$. Considering the high mass ratio, it is traditionally believed that these binary systems will undergo unstable mass transfer followed by CEE. However, if the NS can accrete at a super-Eddington rate and gains mass rapidly, the mass transfer can be stabilized to some extent, especially when the NSs are born massive. We plot the two ULX pulsars in the $P_{\rm orb}$-$L_{\rm acc}$ plane in \autoref{fig:PorbLaccNS} with cyan (M82~X-2) and red (NGC5907~ULX-1) circles. Our results of the six models can roughly reproduce the observed accretion luminosities and orbital periods of M82~X-2, but only Models~\textit{LBCh, HBCh {\rm and} LZCh} can match the observational characteristics of NGC5907~ULX-1.

\subsection{Detection of detached mass-gap BH binaries}

\quad It is impossible to detect BH binaries in X-ray before and after the mass transfer process. To discover BHs in detached binaries, a promising method is based on the radial velocity searches of the optical companions (e.g., \citealt{Giesers+2018,Khokhlov+2018,Thompson+2019,Giesers+2019}). Following previous studies \citep{Breivik+2017,Mashian+2017,Yalinewich+2018,Yamaguchi+2018,Shao+2019+detachedBH}, we evaluate the detectability of detached mass-gap BH binaries.

The apparent magnitude $m_{V}$ of the optical companion in the $V$-band can be obtained by
\begin{equation}
	m_{V}=M_{V}+5\left(2+\log \frac d{\rm kpc}\right)+A_{V},
\end{equation}
where $M_{V}$ is the absolute magnitude of the companion star, $d$ is the distance of the binary from the Sun, and $A_{V}$ is the interstellar extinction in the $V$-band, taken to be $A_{V}=d/{\rm kpc}$ (e.g., \citealt{Yamaguchi+2018}). The distribution of the distance $d$ for each sample is assumed to follow the stellar distribution in the Milky Way \cite[see ][]{Lau+2020}. According to \cite{Bahcall+1980}, we use the following number density ${\mathrm d N}/{\mathrm d l~\mathrm d b~\mathrm d d}$ as a function of the distance from the Galactic center $r=\sqrt{r_0^2+d^2\cos^2b-2d r_0\cos b\cos l}$ in the Galactic plane and the distance perpendicular to the Galactic plane $z=d\sin b$, i.e.,
\begin{equation}\label{eq:d}
	\frac{\mathrm d N}{\mathrm d l~\mathrm d b ~\mathrm d d}\propto d^2\cos b \exp{\left(-\frac{z}{h_z}-\frac{r}{h_r}\right)},
\end{equation}
where $r_0=8.5~{\rm kpc}$ is the distance from the Galactic center to the Sun, $b$ and $l$ are the Galactic latitude and longitude, and $(h_z,h_r)=(0.25~{\rm kpc},3.5~{\rm kpc})$ are the scale lengths for the exponential stellar distributions perpendicular and parallel to the Galactic plane, respectively. In our calculations, we use the Markov chain Monte--Carlo sampler \textsc{emcee} \citep{emcee} to draw the 3-dimensional number density distribution.

The radial velocity semi-amplitude $K$ of the optical companions is
\begin{equation}
	K=\sqrt{\frac{G}{1-e^2}}(M_{\rm c}+M_{\rm BH})^{-1/2}a^{-1/2}M_{\rm c}\sin i,
\end{equation}
where $M_{\rm c}$ and $M_{\rm BH}$ are the masses of the companion and the BH, $e$ and $i$ are the eccentricity and the orbital inclination of the binaries, respectively. In our calculations we take $e=0$ and assume that $i$ follows a uniform distribution in the range of $0-\pi/2$. The astrometric signature $\alpha$ (\citealt{Mashian+2017}) is
\begin{equation}
	\alpha=\frac{M_{\rm BH}}{M_{\rm BH}+M_{\rm c}}\frac{a}{d}.
\end{equation}

\begin{figure*}
	\centering
	\includegraphics[width=\linewidth]{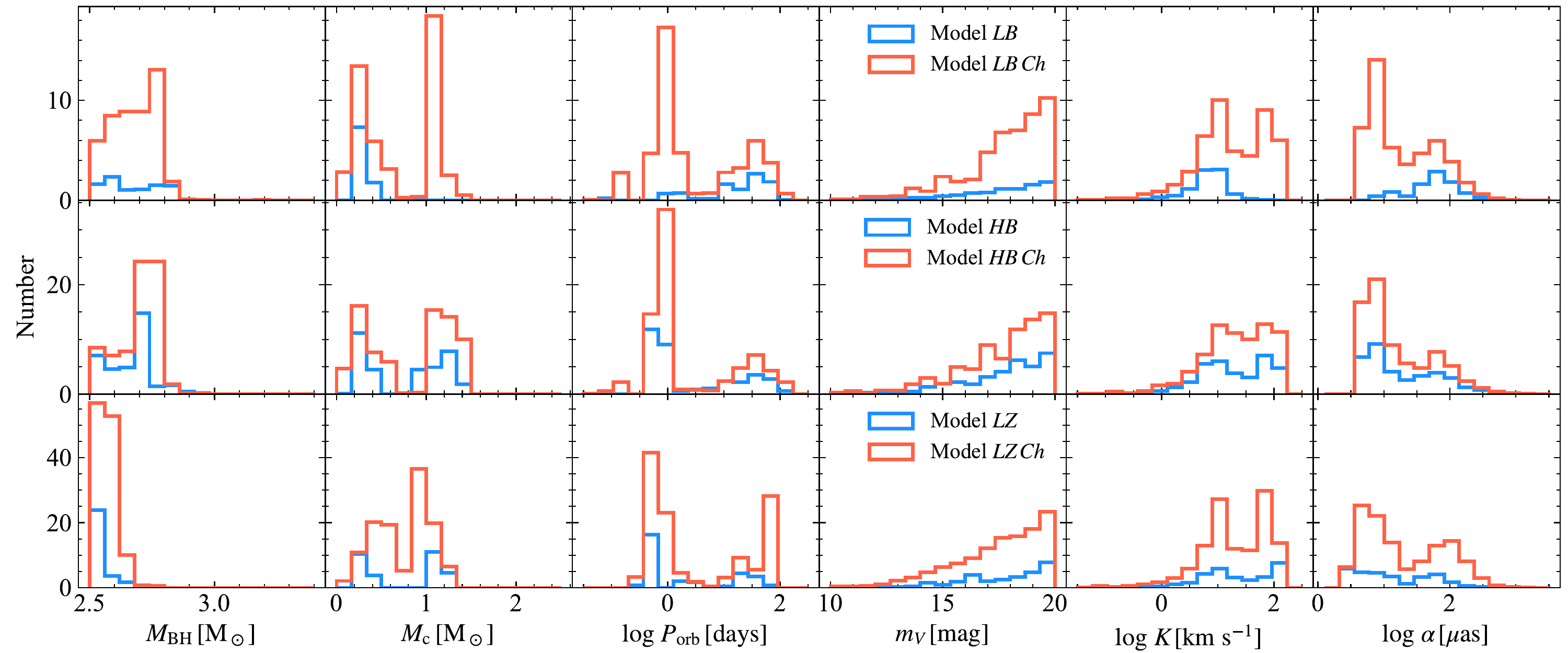}
	\caption{The predicted number distribution of the detached mass-gap BH binaries with optical companions brighter than $20^{\rm m}$ in the Galaxy, as a function of the BH mass, the companion mass, the orbital period, the absolute magnitude, the radial velocity semi-amplitude and the astrometric signature corresponding to the panels from left to right, respectively. The names of the six models in each row are labeled in the fourth column.\label{fig:detached}}

\end{figure*}

In \autoref{fig:detached}, we show the number distribution of detached mass-gap BH binaries with the optical companions brighter than $20^{\rm m}$ in the Galaxy as a function of the BH mass $M_{\rm BH}$, the companion mass $M_{\rm c}$, the orbital period $P_{\rm orb}$, the absolute magnitude $m_{V}$, the radial velocity semi-amplitude $K$, and the astrometric signature $\alpha$. We find that up to dozens to hundreds of detached mass-gap BH binaries may be detected with optical observations. The total number of detached mass-gap BH binaries in each model is summary in \autoref{tab:number}. The companion mass distribution is peaked at $\sim 0.5~{\rm M_\odot}$ and can extend to $\sim 2~{\rm M_\odot}$. The orbital periods display a bimodal distribution peaked at $\lesssim 1~{\rm days}$ and $\lesssim 100~{\rm days}$, which likely originate from the temporary detachment during the XRB phase and the permanent detachment after the mass transfer, respectively. In addition, the distribution of $K$ is roughly peaked at $\sim 100~{\rm km~s^{-1}}$ and $\alpha$ is distributed in the range of $\sim 1-10^3~{\mu\rm as}$, peaked at $\sim 100~{\mu\rm as}$.

\cite{Shao+2019+detachedBH} investigated the formation and evolution of BH binaries born from core collapses of massive stars, and found that the birthrate of incipient BH binaries in the Galaxy is $\sim (4.5-13)\times10^{-5}~{\rm yr^{-1}}$, and there are several hundred detached BH binaries with an optical companion brighter than $20^{\rm m}$. The birthrate of mass-gap BH binaries is $\sim (0.7-3.2)\times10^{-5}~{\rm yr^{-1}}$, implying that BH binaries descended from NS I/LMXBs might make non-negligible contribution to the whole detached BH binary population in the Galaxy.

\subsection{Contribution to GW sources}

\begin{figure*}
	\centering
	\includegraphics[width=0.95\linewidth]{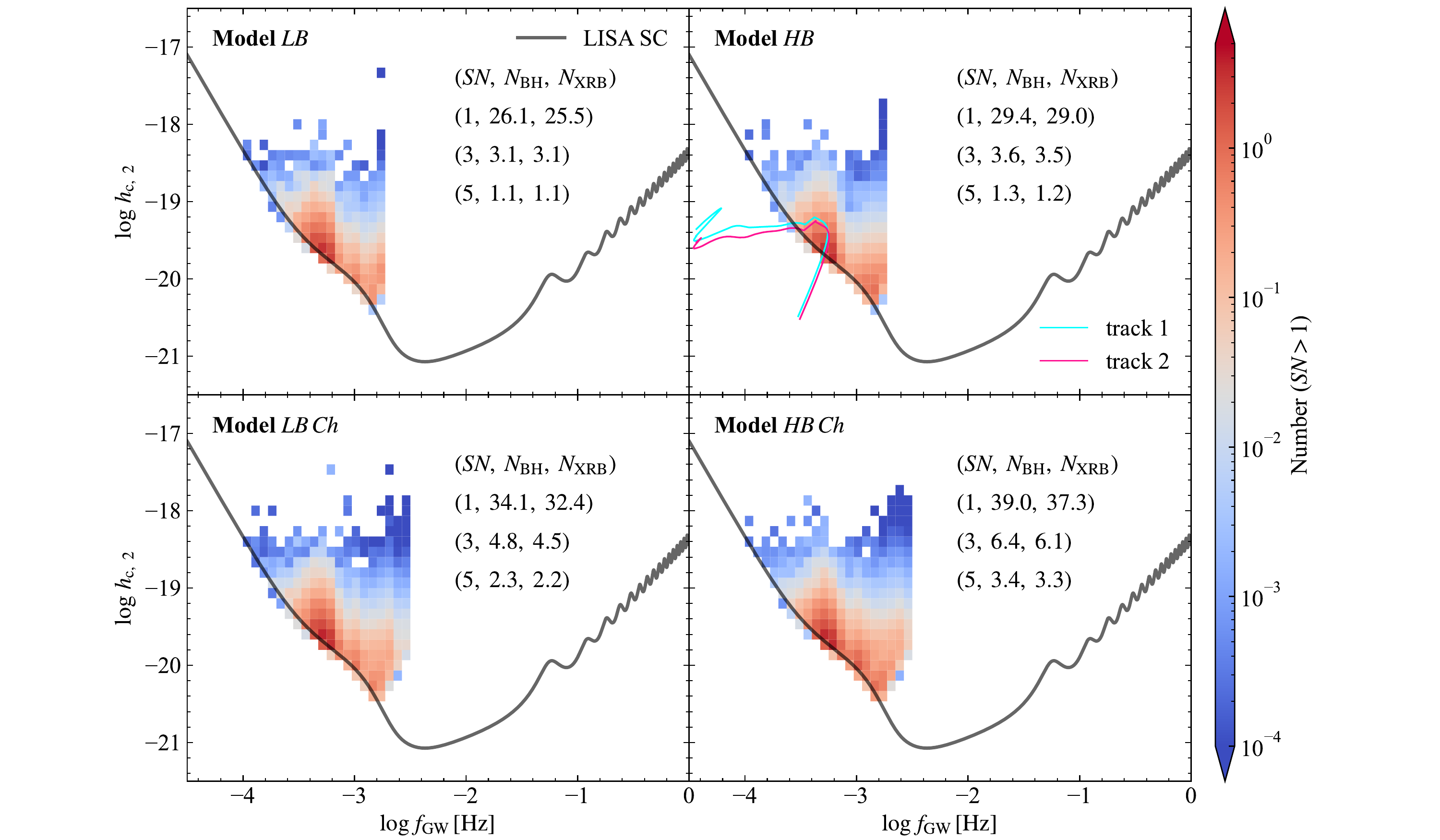}
	\caption{The number distribution of BH binaries that can be potentially detected by \textit{LISA} with $SN>1$ in the GW frequency-characteristic strain ($f_{\rm GW}$-$h_{{\rm c},~2}$) plane. The color in each pixel indicates the number of \textit{LISA} sources. The black solid lines represent the \textit{LISA} sensitivity curve. $N_{\rm BH}$ and $N_{\rm XRB}$ are the numbers of mass-gap BH binaries and BH XRBs that can be detected by \textit{LISA} for a specific $SN$. The cyan and red solid lines in the upper left panel represent the evolutionary pathways of tracks~1 and 2 in the $f_{\rm GW}$-$h_{{\rm c},~2}$ plane at a distance of $0.5~{\rm kpc}$, respectively. The model parameters in each panel are the same as in \autoref{fig:MaccPorb}. Models~\textit{LZ} and \textit{LZCh} do not produce any BH that can be detected by \textit{LISA}, so they do not appear.
	\label{fig:GW}}

\end{figure*}

\begin{figure*}
	\centering
	\includegraphics[width=\linewidth]{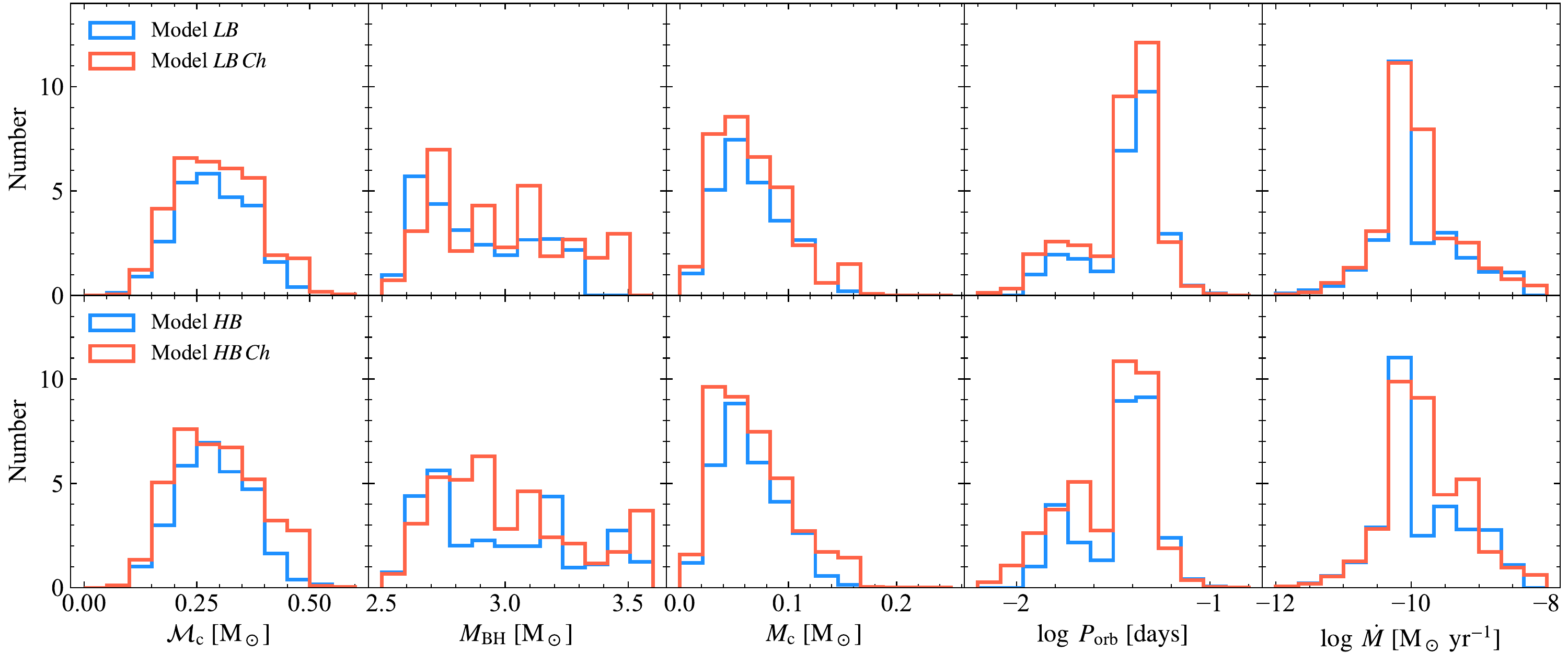}
	\caption{{Number distribution of detectable \textit{LISA} sources with $SN>1$ as a function of the chirp mass, the BH mass, the donor mass, the orbital period and the mass transfer rate corresponding to the panels from left to right, respectively. The names of the models in each row are labeled in the first column.} Models~\textit{LZ} and \textit{LZCh} do not produce any BH that can be detected by \textit{LISA}, so they do not appear.}\label{fig:BHGW}
\end{figure*}

\quad Short-period BH binaries are potential targets of future space-based GW observatories like \textit{LISA} \citep{LISA+2017}. We use the following equation to calculate the characteristic strain $h_{{\rm c},~n}$ at the $n$th harmonic (\citealt{Barack+2004}, see also \citealt{Kremer+2019,Shao+2021}), that is,
\begin{equation}
	h^2_{{\rm c},~n}=\frac{2}{3\pi^{4/3}}\frac{G^{5/3}}{c^3}\frac{\mathcal M_{\rm c}^{5/3}}{d_{\rm L}^2}\frac{1}{f^{1/3}_n}\left(\frac2n\right)^{2/3}\frac{g(n,e)}{F(e)},
\end{equation}
where $\mathcal{M}_{\rm c}=(M_{\rm BH}M_{\rm c})^{3/5}/(M_{\rm BH}+M_{\rm c})^{1/5}$ is the chirp mass of the binary, $d_{\rm L}$ the luminosity distance, $f_{n}=nf_{\rm orb}$ the $n$th harmonics frequency of GWs, $f_{\rm orb}$ the rest-frame orbital frequency, $g(n,e)$ a function of the orbital eccentricity \citep{Peters+1963}, and $F(e)=[1+(73/24)e^2+(37/96)e^4]/(1-e^2)^{7/2}$. We also assume that the BH binaries follow the stellar distribution in the Galaxy and the binary orbits are circular. The peak frequency of the GW is given by $f_{\rm GW}=2(1+e)^{1.1954}(1-e^2)^{-1.5}f_n$ \citep{Wen+2003}, and we take $n=2$ and $f_{\rm GW}=2f_{\rm orb}$. We only select binaries with the signal-to-noise ratio $SN$ larger than unity for the \textit{LISA} mission duration of $T_{\rm LISA}=4~{\rm yr}$ and use the sensitivity curves constructed and fitted by \cite{Robson+2019}. We note that the characteristic strain is reduced by a factor $\min\left[{1,\sqrt{\dot f_{n}T_{\rm LISA}/f_n}}\right]$ to account for the frequency band swept by each source during the observation time $T_{\rm LISA}$ \citep[e.g., ][]{Kremer+2019,Kremer+2020}, where $\dot f_n=n(48/5\pi)(G\mathcal{M}_{\rm c})^{5/3}c^{-5}(2\pi f_{\rm orb})^{11/3}F(e)$ is the derivative of $f_n$ with respect to time.

\autoref{fig:GW} shows the number distribution of BH binaries that can be detected by \textit{LISA} with $SN>1$ in the GW frequency-characteristic strain ($f_{\rm GW}$-$h_{{\rm c},~2}$) plane. The \textit{LISA} sensitivity curve \citep{Robson+2019} is illustrated by the black solid line. The characteristic strain of the detectable sources can reach $\sim 10^{-18}$ with the frequencies $\sim 10^{-4}-10^{-2.5}~{\rm Hz}$. The cyan and red solid lines in the upper-left panel demonstrate the evolutionary paths of tracks~1 and 2, respectively, assuming that the BH binaries are $0.5~{\rm kpc}$ away from the Sun. Binaries like them contribute to almost all of the \textit{LISA} sources.
The number $N_{\rm BH}$ of the detectable mass-gap BH binaries for a specific $SN$ is listed in each panel of \autoref{fig:GW} followed by the number ($N_{\rm XRB}$) of GW-detectable BH XRBs. For $SN>1$, there are $\sim 26-39$ mass-gap BH binaries that can be detected by \textit{LISA} in case~1, while there is no mass-gap BH binaries in case~2. However, the corresponding numbers decrease to $\sim 1-7$ and $0$ with $SN>5$. The initial orbital periods of the incipient NS I/LMXBs that can evolve into mass-gap BH \textit{LISA} sources are in a narrow range of $\sim 0.5-2~{\rm days}$, depending on the bifurcation period.

\autoref{fig:BHGW} shows the number distribution of the mass-gap BH binaries that can be detected by \textit{LISA} with $SN>1$ as a function of the chirp mass $\mathcal{M}_{\rm c}$, the BH mass $M_{\rm BH}$, the companion mass $M_{\rm c}$, the orbital period $P_{\rm orb}$ and the mass transfer rate $\dot M$ (from left to right panels). There is no detectable BH binary with $SN=1$ in Models~\textit{LZ {\rm and} LZCh} with metallicity $Z=0.001$. The BH masses and orbital periods cover the range of $\sim 2.5-3.5~{\rm M_\odot}$, $\sim 0.01-0.1~{\rm days}$, respectively. Almost all of the \textit{LISA}-detectable BH binaries are transient X-ray sources with mass transfer rate $\sim 10^{-12}-10^{-8}~{\rm M_\odot~yr^{-1}}$ and orbital period $P_{\rm orb}\sim 0.01-0.1~{\rm days}$ (see also \autoref{fig:PorbMdotBH}). The chirp mass of the mass-gap BH sources varies in the range of $\sim0.1-0.5~{\rm M_\odot}$ and is peaked at $\sim 0.25~{\rm M_\odot}$. In this respect, mass-gap BH binaries may be discriminated from BH binaries formed from core-collapse SNe, in which $\mathcal{M}_{\rm c}\sim 0.8-2~{\rm M_\odot}$, $\sim 1.2-4~{\rm M_\odot}$, and $\sim 3-12~{\rm M_\odot}$ for BH+WD binaries, BH+NS binaries, and BH+BH binaries \citep{Shao+2021}.

\section{Discussion and Summary}\label{section:dis&sum}

\begin{figure}
    \centering
    \includegraphics[width=\linewidth]{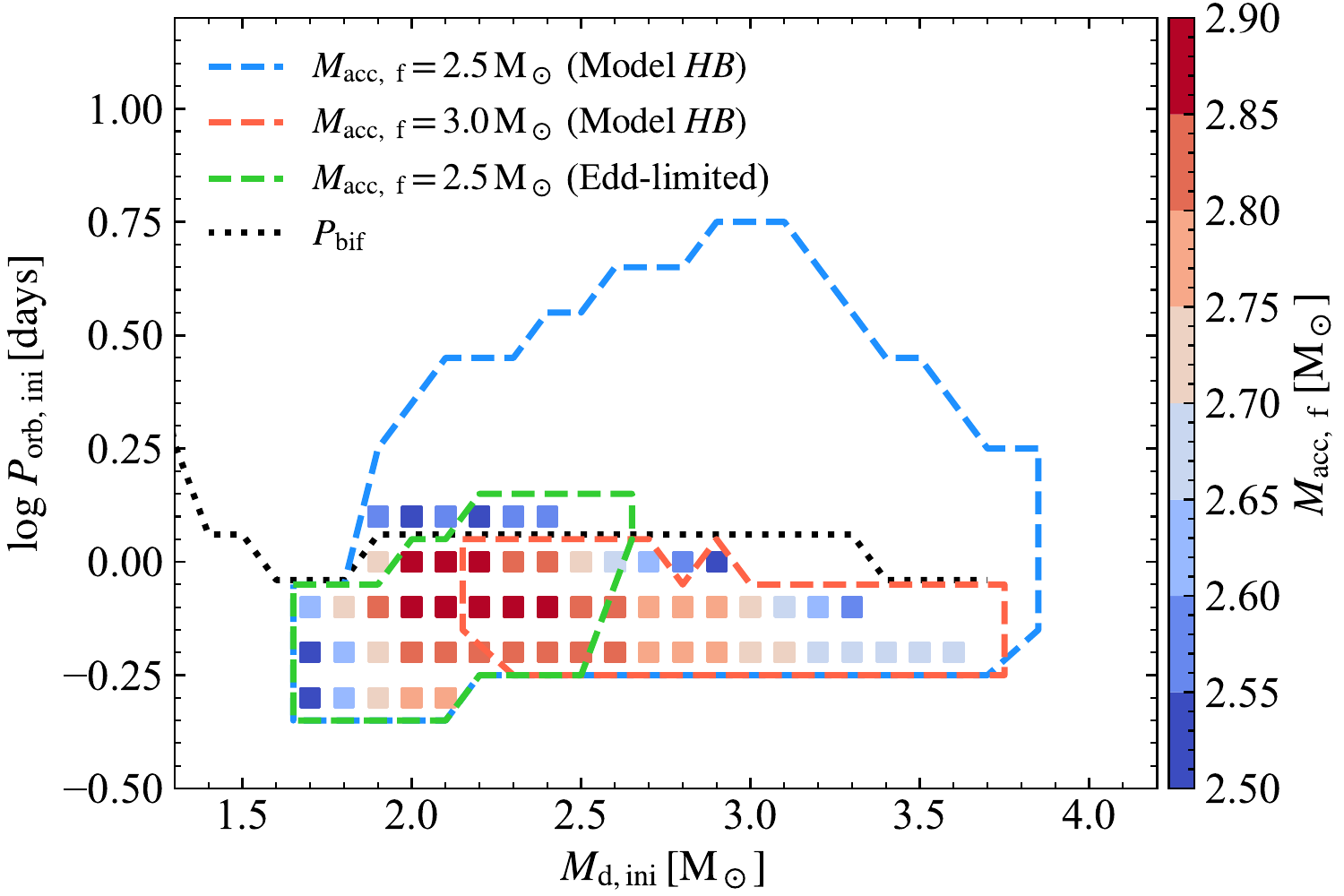}
    \caption{The colored squares demonstrate the final masses of the accretors in the $M_{\rm d,~ini}$-$P_{\rm orb,~ini}$ plane with magnetic field decay being taken into account. The blue and the red dashed curves indicate $M_{\rm acc,~f}=2.5~{\rm M_\odot}$ and $3.0~{\rm M_\odot}$ in Model~\textit{HB} respectively, and the green dashed curve indicates $M_{\rm acc,~f}=2.5~{\rm M_\odot}$ with the Eddington-limited accretion prescription. The black dotted line shows the bifurcation period $P_{\rm bif}$ as a function of $M_{\rm d,~ini}$.}\label{fig:decay}
\end{figure}

\begin{figure}
    \centering
    \includegraphics[width=\linewidth]{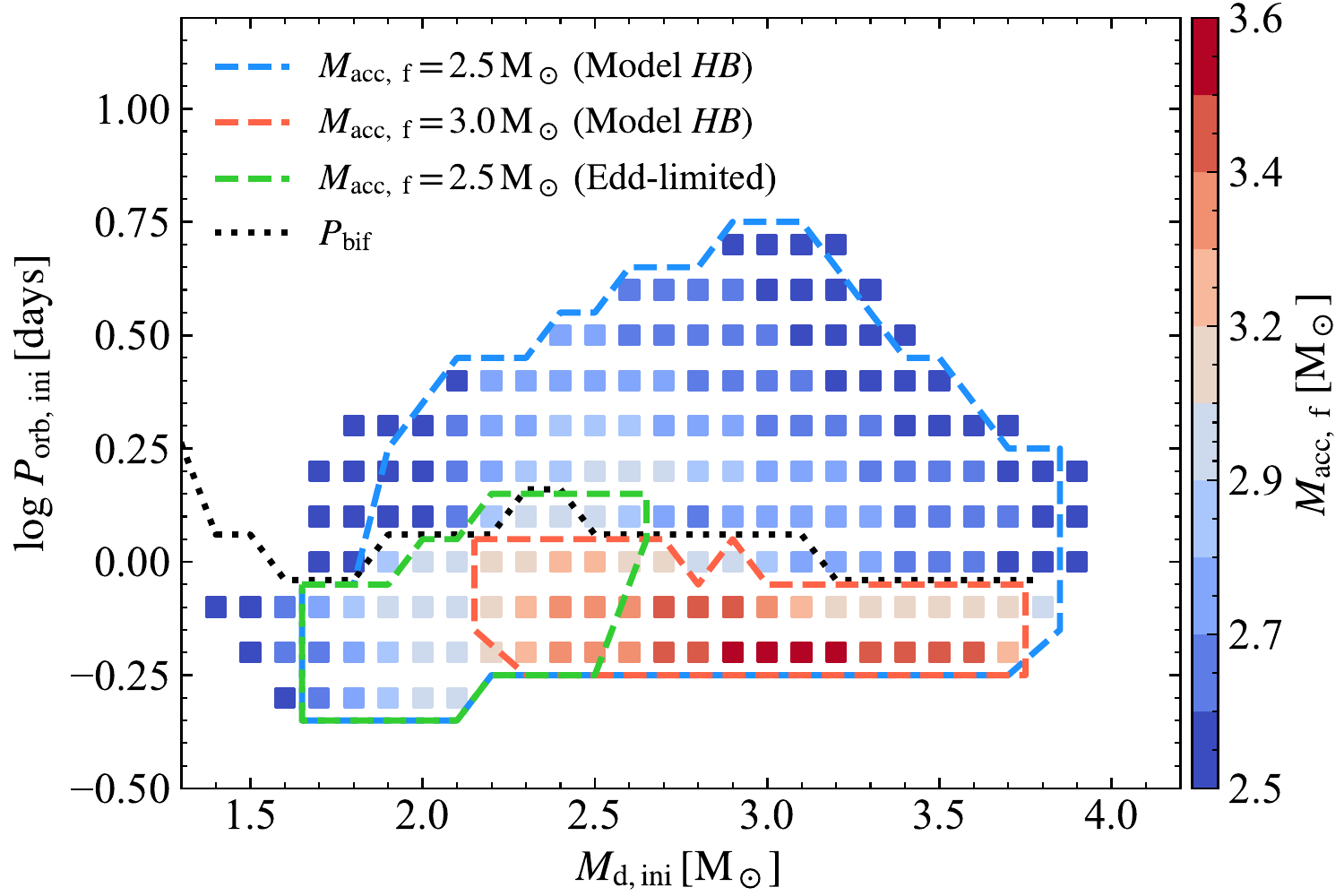}
    \caption{Similar to \autoref{fig:decay}, the colors of the squares represent the final masses of BHs with $\beta_{\rm mt}=0.5$ for transient NSs.}
    \label{fig:beta}
\end{figure}

\quad In this work, we explore the possible formation of mass-gap BHs from super-Eddington accreting NSs. In the traditional picture, the growth of the NS mass is limited by the Eddington accretion rate. Our work is motivated by two facts. (1) In the observational aspect, the luminosities of some ULX pulsars have been inferred to be $\gtrsim 10^{41}~{\rm erg~s^{-1}}$, and their sinusoidal pulse profiles imply that strong beaming of the X-ray radiation is not favored. Thus, the observed (isotropic) luminosities are likely close to their true accretion luminosities and the NSs are accreting at a super-Eddington rate. (2) In the theoretical aspect, it is difficult for a single SN mechanism account for both the mass gap and the existence of a few low-mass BHs. Thus, AIC of NSs in I/LMXBs may provide an alternative formation channel of mass-gap BHs like GRO J0422+32.

Our calculations show that mass-gap BHs descending from NS I/LMXBs generally experience the evolutionary stages of detached, mass transferring, and merging binaries, and can be potentially detected with optical, X-ray and GW methods, respectively. The majority of mass-gap BH LMXBs are transient sources (see \autoref{fig:PorbMdotBH}), similar to traditional BH LMXBs. Our calculations show that there might be tens to hundreds of detached mass-gap BH binaries with optical companions brighter than $20^{\rm m}$, which can be observed by future optical observations. GW radiation presents another useful way to detect the mass-gap BHs. We predict that there might be up to 3 BH binaries to be detected by \textit{LISA} with $SN>5$. Their chirp mass distribution may provide unique feature to test the AIC formations channel for mass-gap BHs.

Our results are subject to the uncertainties related to accretion physics, like accretion-induced magnetic field decay and the accretion efficiency of transient NS XRBs. Below we discuss their influence separately.

The apparent correlation between the mass transferred mass and the magnetic field strength in LMXBs suggests that NS magnetic fields may decrease due to accretion, so the critical mass accretion rate in \autoref{eq:mdotcch} would change with time accordingly. 
Here we follow \cite{Osowski+2011} to adopt an empirical form of accretion-induced magnetic field decay,
\begin{equation}\label{eq:B_decay}
    B=(B_0-B_{\rm min})\exp{
\left(-\frac{\delta M}{\Delta M_{\rm dec}}\right)}+B_{\rm min},
\end{equation}
where $B_0$ is the initial magnetic field, $B_{\rm min}$ the minimal (bottom) magnetic field of the NS, $\delta M$ the acrreted mass of the NS, and $\Delta M_{\rm dec}$ is a constant. To quantify the effect of magnetic field decay, we consider a model with $B_{\rm min}=5\times10^{8}~{\rm G}$ and $\Delta M_{\rm dec}=0.05~{\rm M_\odot}$, and other parameters are the same as in Model~\textit{HB}. When the magnetic field becomes sufficiently low, $R_0$ decreases to be comparable with $R_{\rm NS}$. So we take the critical accretion rate to be $\max[\dot M_{\rm crit},\dot M_{\rm Edd}]$. In \autoref{fig:decay}, the final masses of the accretors are depicted with the colored squares in the $M_{\rm d,~ini}$-$P_{\rm orb,~ini}$ plane. For comparison, the blue and the red dashed curves demonstrate the binary distribution with $M_{\rm acc,~f}=2.5~{\rm M_\odot}$ and $3.0~{\rm M_\odot}$ in Model~\textit{HB} respectively, and the green dashed curve indicates $M_{\rm acc,~f}=2.5~{\rm M_\odot}$ with the Eddington-limited accretion prescription. When magnetic field decay is considered, the allowed parameter space for the formation mass-gap BHs is significantly smaller than that in Model~\textit{HB}, and the BH binaries are mainly formed from relatively compact (and converging) systems ($P_{\rm orb,~ini}\lesssim 1.3$ days). In this case, the maximum mass of the BHs is about $2.89~{\rm M_\odot}$. The birth rate of the BH binaries and the total number of BH XRBs in the Galaxy reduce from $1.39\times10^{-5}~{\rm yr^{-1}}$ to $3.29\times10^{-6}~{\rm yr^{-1}}$ and from $2.3\times10^4$ to $1.66\times10^4$, respectively. However, the number of the detectable \textit{LISA} sources with $S/N>5$ remains unchanged ($\sim 1$), because they result from converging binaries. If we take smaller value for $\Delta M_{\rm dec}$, the parameter space will get further smaller and be close to the Eddington-limited accretion case (the green dashed curve in \autoref{fig:decay}).

Next we consider the accretion efficiency from an unstable disk. Here the main uncertainty is how much mass an NS can accrete during outbursts, which depends on the maximum mass transfer rate $\dot M_{\rm max}$ and the $\dot M$ profile during the outbursts, as well as the spin period and magnetic field of the NS. \cite{Bhattacharyya+2017} computed the evolution of accreting NSs in transient LMXBs through a series of outburst and quiescent phases, assuming a simplified, linear $\dot M$ profile during both outburst rise and decay. They found that the ratio of the equilibrium spin periods of persistent to transient NSs with the same $\dot M$ is around $2-6$ \citep[see also,][]{Bhattacharyya+2021}. From this we can infer a rough estimate of the accretion efficiency $\sim 0.5$. We then calculate a model similar to Model~\textit{HB} but with $\beta_{\rm mt}=0.5$ for transient NS XRBs. \autoref{fig:beta} shows the distribution in the $M_{\rm d,~ini}$-$P_{\rm orb,~ini}$ plane, and the final masses of the BHs are depicted with colored squares. Compared with Model~\textit{HB} (the blue dashed curve), the allowed parameter space for the formation mass-gap BHs varies slightly. The birth rate of BH binaries and the number of BH XRBs increase by a factor $\lesssim 20\%$. The number of detectable \textit{LISA} sources with $SN>5$ is almost unchanged and is still $\sim 1$. The reason is that DI usually occurs when the mass transfer rate is $\lesssim 10^{-9}~{\rm M_\odot~yr^{-1}}$ for NS accretors, so its influence on angular momentum loss and mass growth of the accretor is insignificant.

We conclude that our results shown in Figures~\ref{fig:MaccPorb}--\ref{fig:BHGW} represent the optimistic case for the formation of mass-gap BH from AIC of NSs, and the predicted birthrate and total number of BH binaries in Tables~\ref{tab:br} and \ref{tab:number} should be regarded as the upper limits. Nevertheless, our study shows that super-Eddington accretion may have non-negligible influence on the evolution of NS I/LMXBs.

\section*{Acknowledgments}
\quad We are grateful to an anonymous referee for helpful comments. This work was supported by the National Key Research and Development Program of China (2021YFA0718500), the Natural Science Foundation of China under grant No. 12041301, 12121003, and Project U1838201 supported by NSFC and CAS. The computation was made by using the facilities at the High-Performance Computing Center of Collaborative Innovation Center of Advanced Microstructures (Nanjing University).

\section*{Data Availability}
\quad All data underlying this article will be shared on reasonable request to the corresponding author.

\bibliographystyle{mnras}
\bibliography{references.bib}

% Don't change these lines
\bsp	% typesetting comment
\label{lastpage}
\end{document}